\documentclass[aps,pre,floatfix,twocolumn,nofootinbib,showpacs]{revtex4}
\usepackage{dcolumn}
\usepackage{bm}
\usepackage[dvips]{epsfig}
\usepackage{graphicx} \usepackage{amsmath} \usepackage{amssymb}
\newcommand{\comment}[1]{}
\newcommand{\BEQ}{\begin{equation}}
\newcommand{\EEQ}{\end{equation}}
\newcommand{\BEA}{\begin{eqnarray}}
\newcommand{\EEA}{\end{eqnarray}}
\renewcommand{\d}{{\rm d}}

\renewcommand{\L}{{\rm L}}
\newcommand{\R}{{\rm R}}
\newcommand{\J}{\bar{J}}
\renewcommand{\k}{{\rm k}}
\newcommand{\E}{{\bf {\cal E}}}
\newcommand{\F}{{\bf F}}

\renewcommand{\S}{{\bf S}}
\newcommand{\p}{\bar{p}}
\renewcommand{\P}{{\cal P}}
\newcommand{\x}{\hat{x}}
\renewcommand{\bm}{\bar{\mu}}
\renewcommand{\a}{\alpha}
\renewcommand{\b}{\beta}

\begin{document}

\title{Adaptive machine and its thermodynamic costs}

\author{Armen E. Allahverdyan$^{1,2)}$ and Q. A. Wang$^{1)}$}

\address{$^{1)}$Laboratoire de Physique Statistique et Syst\`emes Complexes,
ISMANS, 44 ave. Bartholdi, 72000 Le Mans, France\\
$^{2)}$Yerevan Physics Institute, Alikhanian Brothers Street 2, Yerevan 375036, Armenia }

\begin{abstract} We study the minimal thermodynamically consistent model
for an adaptive machine that transfers particles from a higher chemical
potential reservoir to a lower one. This model describes essentials of
the inhomogeneous catalysis.  It is supposed to function with the
maximal current under uncertain chemical potentials: if they change, the
machine tunes its own structure fitting it to the maximal current under
new conditions. This adaptation is possible under two limitations. {\it
i)} The degree of freedom that controls the machine's structure has to
have a stored energy (described via a negative temperature). The origin
of this result is traced back to the Le Chatelier principle.  {\it ii)}
The machine has to malfunction at a constant environment due to
structural fluctuations, whose relative magnitude is controlled solely
by the stored energy. We argue that several features of the adaptive
machine are similar to those of living organisms (energy storage,
aging). 
\end{abstract}


\comment{
}

\maketitle

\section{Introduction} 

Adaptation is one of the paradigms of biology and complex systems
theory, but its investigations \cite{ho,crowley,tyukin,formal,taivo}
rarely start from the first principles of thermal physics (instead, they
proceed with mathematical \cite{tyukin} or qualitative approaches
\cite{formal,taivo}). Hence not much is known about the physical costs
of adaptation. Besides its fundamental importance, this question is
relevant due to increasing interest in smart (self-controling) materials
\cite{healing} and due to miniaturization of technologies that make the
external control impossible or obsolete. 

Consider a machine that transports matter with the maximal current
allowed by external constraints.  This optimal functioning will be seen
to demand a good fit between its structure and external environment
(making the machine somewhat similar to an organism). For exploiting
such a machine in an uncertain environment one can control it
externally, or design a specific direct interaction between the
environment and the structure. Here we explore the most interesting
possibility: upon environmental changes, the machine tunes its own
structure so as to work optimally under new environment. Such a machine
is {\it adaptive} without external control. Ordinary macroscopic
machines are not adaptive in this sense: their structure is either
predetermined or is controlled externally. This is why the laws of
thermodynamics focus on the impossibility of achieving certain tasks via
external fields without feedback \cite{lindblad} \footnote{Feedback
control, where the action of external fields takes into account some
information on the system's state, is also a traditional subject of
thermodynamics \cite{maxwell,fee,fee1,seifert_review}. However, in the majority of
papers on this subject people are interested by the usage of feedback in
extracting more work or in reducing the entropy \cite{fee1,seifert_review}; two
original intentions of the Maxwell's demon \cite{maxwell} (see, however,
\cite{fee}). Moreover, the feedback is typically described
externally, i.e. without making the controller an integral part of the
described set-up. In contrast, we are interested here by the feedback
processes that maintain a maximal current on the face of environmental
changes, and we describe the controller explicitly. }. 

But small machines are able to alter their own structure. In certain
enzymes and ion channels the functional part (performing the catalysis)
couples to the conformational part that in its
turn back-reacts on the functional part \cite{blum,conformon}. 
Structure-function interaction exists also in 
inhomogeneous catalysis \cite{catal}. It modifies the 
catalyst's structure and changes the catalytic current.

Our purpose is to understand thermodynamic limits of adaptation via the
{\it minimal} model of a structure-adaptive machine transporting
particles from one reservoir to another. We choose this model for three
reasons. First, it realizes the simplest and most fundamental
machine-like function (catching and releasing); hence its understanding
can influence the design of future adaptive machines. Second, the model
adequately describes the essentials of inhomogeneous catalysis. Third,
this is a step towards studying more complex systems of biochemical
catalysis which evolved to increase their current \cite{ho,crowley}.
Our personal motivation of studying adaptive machines is a
belief that these non-living systems may demonstrate certain key
features of living organisms. 

Here is a brief description of the adaptive machine to be elaborated
below. For an environment with given chemical potentials, our model
machine transports particles from the higher chemical potential to the
lower one and does so with the maximal current (or speed) once its
structure (energies of its states) fits that particular environment.
Upon changing the chemical potentials the functional part tunes the
structure so that the machine functions with the maximal current under
new conditions. We study thermodynamical costs of this adaptation. (We
focus on changes of chemical potentials, since the chemical potential
difference is the driving force of the particle transport.)

This work is organized as follows. The next two sections define the main
ingredients of the model.  Section \ref{adaptation} defines the concept
of adaptation, as applied to our situation.  Sections \ref{adaptation}
and \ref{flucto} identify the main thermodynamic costs of adaptation.
Section \ref{alternative} discusses certain alternative set-ups of
adaptation, e.g. when the controlling degree of freedom is allowed to
sense directly the uncertain environment. The last section summarizes
our work. Here we also discuss how the thermodynamic costs of adaptation
relate to basic characteristics of aging and energy storage known for
living organisms. The paper has five appendices. 

\comment{The functional and structural parts
will be modelled via two degrees of freedom coupled to each other and
forming together an autonomous system. The necessary separation between
them is then achieved due to the time-scale difference (functional part
is fast, while the structural one is slow) and the fact that the
structural part does not interact directly with the environment. }

\section{The model: functional degree of freedom} 
\label{functional}

\begin{figure}
\includegraphics[width=10cm]{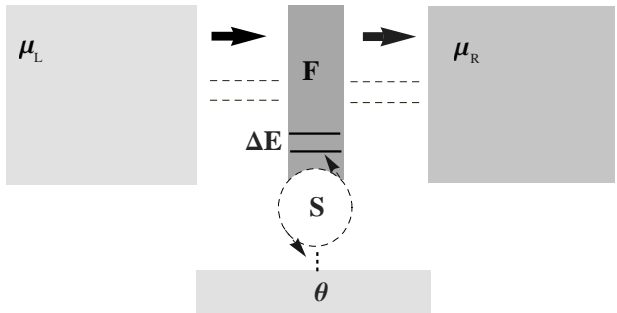}
\vspace{-0.8cm}
\caption{ The global scheme of the model. Two-level system $\F$ transfers 
particles from a reservoir with chemical potential $\mu_\L$ to 
that with $\mu_\R$. The controller $\S$ tunes the energy difference 
$\Delta E$ of $\F$ and interacts with a bath at temperature $1/\theta$.
} 
\label{fig2}
\end{figure}

Our model has two degrees of freedom: functional and structural; see
Fig.~\ref{fig2}. They couple to each other forming together an
autonomous system.  First we shall discuss the dynamics of the
functional degree of freedom $\F$ assuming that the structural degree of
freedom is fixed. 

\subsection{Definition of $\F$}
\label{defo}

$\F$ is a simple model for a trap (or adsorption center). It has two
states $\F_i$ ($i=1,2$): an empty state $\F_1$ whose energy is $E_1$ and
a filled state $\F_2$ with one particle and energy $E_2$.  Hence each
state has energy $E_i$ and carries $N_i$ particles($i=1,2$): $N_2=1$ and
$N_1=0$; see Fig.~\ref{fig1}. Without coupling to external reservoir(s), $\F$ will stay
indefinitely in one of its states, since the energy and the particle
number are conserved.  Hence any change between the states of $\F$ is
driven externally. 

Let us first assume that $\F$ interacts with an equilibrium reservoir at
chemical potential $\mu$ and temperature $T=1$ (this value of
temperature is chosen conventionally, since temperature gradients will
not play any role in our study). The stationary (time-independent)
probabilities $\p_i^{\rm [eq]}$ of $\F_i$ ($i=1,2$) have the
grand-canonical (equilibrium) Gibbsian form \cite{landau,kampen}
\footnote{In this paper we study distinguishable classical particles, while discrete
energy states have the usual (e.g. in chemical physics)
meaning of deep minima of potential energy. But we stress that the probability $\p_i^{\rm
[eq]}\propto e^{-(E_i+\mu N_i)/T}$ for a $n$-state system ($n$ may be
infinite) interacting with an equilibrium reservoir at temperature $T$
and chemical potential $\mu$ is the general expression for the grand-canonical equilibrium \cite{landau}.
Here each state $i$ has energy $E_i$ and the particle number $N_i$. For
instance, this description applies to indistinguishable Bose or Fermi
particles \cite{landau}. We do not consider these cases in the present article, but
for illustrative purposes let us remind how this formula applies to
non-interacting Fermi particles; each particle has energy levels
$\varepsilon_1,..., \varepsilon_M$, and not more than one particle can
be in the same energy level. Now the states $i$ can be parametrized via
the filling numbers $n_k=0,1$ ($k=1,...,M$) so that $n_k$ is the number of
Fermi particles having the energy $\varepsilon_k$. Now $E_i=E\{n_k\} =\sum_{k=1}^M \varepsilon_kn_k$,
$N_i=N\{n_k\} =\sum_{k=1}^M n_k$ and $\p^{\rm
[eq]}\{n_k\}= e^{-(E\{n_k\}+\mu N\{n_k\})/T}/Z$, where $Z=\prod_{k=1}^M (1+e^{-(\varepsilon_k-\mu)/T})$
is the statistical sum for Fermi particles.
}
\BEA
\label{budu}
\p_i^{\rm [eq]}\propto e^{-E_i+\mu N_i}. 
\EEA
The relaxation of $\F$ towards its equilibrium state (\ref{budu}) can be described by the master equation
\cite{kampen}:
\BEA
\label{1}
\dot{p}_i\equiv {\d p_i}/{\d t}={\sum}_{j}[\rho_{ij}p_j-\rho_{ji}p_i], \qquad i,j=1,2,
\EEA
where $p_i$ ($p_1+p_2=1$) is the probability of $\F_i$ and $\rho_{i\not
=j}$ is the transition rate $\F_j\to\F_i$.  Since the bath is in
equilibrium, $\rho_{i\not =j}$ hold the detailed balance condition:
\BEA
\label{2.0}
&&\rho_{12}=e^{(E_2-E_1)-\mu (N_2-N_1)}\rho_{21}=
e^{\Delta E-\mu}\rho_{21}, \\
&&\Delta E\equiv E_2-E_1.
\label{oro}
\EEA
Eq.~(\ref{2.0}) ensures that the stationary state of (\ref{1}) coincides with (\ref{budu}).

Without loss of generality we parametrize transition rates $\rho_{12}$ and $\rho_{21}$ as
\BEA
\label{66.00}
\rho_{12}=\frac{1}{\tau}e^{\frac{1}{2}[\Delta E-\mu]}, ~~
\rho_{21}=\frac{1}{\tau}e^{\frac{1}{2}[-\Delta E+\mu]},
\EEA
where $\tau$ is the time-scale induced by the interaction with the
reservoir.  In (\ref{1}), $\tau$ scales the running time (and hence the
relaxation time), but does not appear in the equilibrium probabilities
(\ref{budu}). Microscopic derivations of the master equation show that
$\tau$ depends on the features of the reservoir (e.g. its energy
spectrum), but can also depend on the internal parameter $\Delta
Q=\Delta E-\mu$; see \cite{weiss} and (\ref{sta}) below. Since $\Delta
Q$ is the heat received or transferred to the reservoir, the time-scale
$\tau$ has the global minimum at $\Delta Q=0$: it takes longer to
transfer a larger amount of heat.  This holds for all physical cases we
are aware of.

\subsection{Master equation for two reservoirs}
\label{masto}

In the equilibrium state (\ref{budu}) all currents nullify; this is the
main message of the equilibrium state and it is ensured by the detailed
balance condition (\ref{2.0}) \cite{kampen}. 

We are interested by transport due to a chemical potential gradient. Hence 
we assume that $\F$ simultaneously couples with two equilibrium reservoirs (L and R) of
energy and particles \footnote{A somewhat more realistic assumptions
would be that $\F$ consists of two interacting two-level systems $\F^{[1]}$
and $\F^{[2]}$ so that $\F^{[1]}$ ($\F^{[2]}$) couples only with L (R). If the
interaction between $\F^{[1]}$ and $\F^{[2]}$ is very strong|they are forced to
be simultaneously in their up or down states|we can effectively replace
$\F^{[1]}$ and $\F^{[2]}$ by a single two level system that couples simultaneously 
with two thermal baths; the states, where $\F^{[1]}$
is up while $\F^{[2]}$ is down (or {\it vice versa}) have too large energies
to be populated.}. 
Their temperatures are equal,
\BEA
\label{staro}
T_{\rm L}=T_{\rm R}=T=1, 
\EEA
but the chemical potentials are different 
\BEA
\mu_{\rm L}>\mu_{\rm R}. 
\EEA
$\F$ will transport particles from L to R; see Fig.~\ref{fig1} and
(\ref{kati}) below. Its dynamics is described by a master equation
(\ref{1}), but now once $\F$ couples simultaneously with L and R,
\BEA
\label{simsim}
\rho_{ij}=\rho_{ij|\L}+\rho_{ij|\R}, 
\EEA
where $\rho_{ij|\L}$ and $\rho_{ij|\R}$ are the 
transition rates driven by separate reservoirs \cite{rozenbaum}. 
Since $\L$ and $\R$ are in equilibrium, $\rho_{ij|\k}$ satisfy detailed balance [cf. (\ref{2.0})]
\BEA
\label{2}
\rho_{12|\k}=e^{\Delta E-\mu_{\k}}\rho_{21|\k}, \qquad \k=\L,\R.
\EEA
As (\ref{1}) shows, $\F$ relaxes in time from any initial probability to
the stationary (but generally non-equilibrium) probability 
\BEA
\label{stati}
\p_i={\rho_{ij}}/{(\rho_{ij}+\rho_{ji})}.
\EEA
If L and R are in mutual equilibrium ($\mu=\mu_{\L}=\mu_{\R}$), $\p_i$ reverts 
via (\ref{2}) to the equilibrium (Gibbs) probability (\ref{budu}).
We can apply the same parametrization as in (\ref{66.00})
\BEA
\label{6.0}
\rho_{12|\k}=\frac{1}{\tau_\k}e^{\frac{1}{2}[\Delta E-\mu_\k\Delta N]}, ~~
\rho_{21|\k}=\frac{1}{\tau_\k}e^{\frac{1}{2}[-\Delta E+\mu_\k\Delta N]},
\EEA
where $\tau_\k$ is the time-scale of $\F-\k$ interaction ($\k=\L,\R$).
The discussion given after (\ref{66.00}) now applies to $\tau_\k$ in 
separate. In particular, $\tau_\k$|as a function of $\Delta_\k
Q=\Delta E-\mu_\k$ (heat received or transferred to the reservoir $\k$)|has
the global minimum at $\Delta Q_\k=0$.

In contrast to the equilibrium situation, now the time-scales $\tau_\k$
will generally appear also in the stationary probability (\ref{stati});
see (\ref{7.77}) below.

\begin{figure}
\includegraphics[width=7cm]{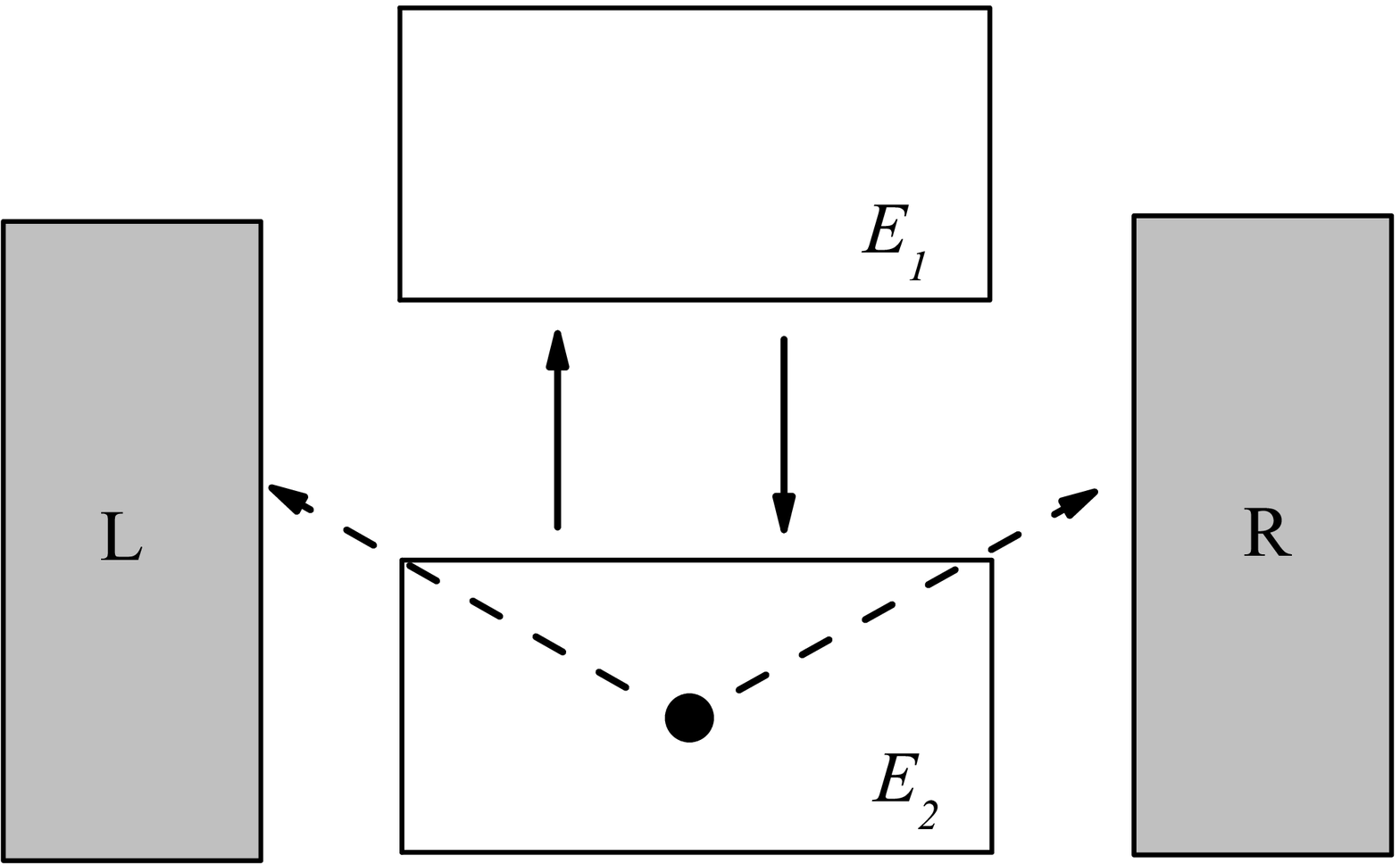}
\vspace{0.8cm}
\caption{ 
Two states of the functional degree of freedom $\F$; see section
\ref{defo} and \ref{masto}.  The state $\F_1$ ($\F_2$) is empty (filled
with one particle) and has energy $E_1$ ($E_2$).  The particle in $\F_2$
can move to one of the reservoirs (L or R) causing transition $\F_2\to
\F_1$. The reverse transition $\F_1\to \F_2$ is due to catching a
particle from one of the reservoirs. The mechanism of transporting
particles from L to R is that a particle comes from L and then jumps to R; see (\ref{kati}). 
} 
\label{fig1}
\end{figure}

\subsection{Particle current}

\subsubsection{General definition}

Our main target is the particle current: the 
mean number of particles entering to $\F$ per time-unit from
the k-reservoir. To find the current, let us note that the time-derivative
of the average number of particles 
\BEA
\label{dubinushka}
&&\frac{\d}{\d t}{\sum}_{i}N_ip_i={\sum}_{i}N_i\dot{p}_i ={\sum}_{\rm k=L,R} J_{\rm k},\\
&&J_{\rm k}={\sum}_{ij}N_i[\rho_{ij|\k}\,p_j-\rho_{ji|\k}\,p_i], ~ \k=\L,\R,
\label{barbi}
\EEA
is a sum $J_{\rm L}+J_{\rm R}$ of two separate contributions. $J_{\rm k}$ is due to
interaction with the reservoir $\k$, i.e. it is formed by the transition
rates $\rho_{ij|\k}$ coming from the reservoir $\k$. These transitions rates satisfy
the detailed balance with respect to the reservoir $\k$; see
(\ref{simsim}, \ref{2}). Since the particle number is conserved during the interactio of $\F$ with each reservoir, the
contribution $J_\k$ to $\frac{\d}{\d t}{\sum}_{i}N_ip_i$ coming from the
reservoir $\k$ is identified with the current of particles from the
reservoir $\k$ \cite{seifert_review,spohn} \footnote{This widely applied identification of the particle current becomes 
completely explicit within microscopically derived models of master-equations; see \cite{weiss,schaller} for reviews.}.
Then (\ref{dubinushka}) expresses the conservation law
\footnote{If there is only one reservoir acting
on $\F$, the particle current from it can be measured simply via the
time-derivative $\frac{\d}{\d t}{\sum}_{i}N_ip_i$ of the average number
of particles. This, of course, already implies that the number of
particles is conserved. If there are two or more reservoir acting on
$\F$, the current of particles coming from a speific reservoir has to be
measured via the reservoir.}.

Eqs.~(\ref{1}, \ref{barbi}) lead to
\BEA
\label{4.0}
J_{\rm k}=\rho_{21|\k}\,p_1-\rho_{12|\k}\,p_2, \qquad \k=\L,\R,
\EEA
where we recalled that $N_2=1$ and $N_1=0$. It should be intuitively
clear from (\ref{4.0}) that $J_{\rm k}$ is indeed the average number of
particles entering from the reservoir $\k$: $\rho_{21|\k}\,p_1$ is
(proportional to) the probability for $\F$ to make transition
$\F_1\to\F_2$ induced by the reservoir $\k$ (hence $\F$ catches a particle
from the reservoir $\k$). From this one subtracts the probability
$\rho_{12|\k}\,p_2$ of the reverse event: the particle leaving from $\F$
to the reservoir $\k$. 

\subsubsection{Stationary situation}

In the stationary (time-independent) situation $\J_\L+\J_\R=\frac{\d
}{\d t}\sum_iN_i\p_i=0$; see (\ref{dubinushka}, \ref{barbi}). As
expected, there is only one independent current in the stationary
situation. Using (\ref{stati}, \ref{4.0}) we get
\BEA
\J_\L=\frac{\rho_{21|\L}\rho_{12|\R}-\rho_{12|\L}\rho_{21|\R}}{\rho_{12}+\rho_{21}}.
\label{kati}
\EEA
Eq.~(\ref{kati}) has a transparent interpretation that explains the
mechanism of the particle transport from L to R; see also
Fig.~\ref{fig1} in this context. Indeed, $\rho_{21|\L}$ is (proportional
to) the probability that $\F$ transits $\F_1\to \F_2$ under influence of
L. This means that the particle in $\F$ came from L. Likewise,
$\rho_{12|\R}$ is the probability that the particle will leave to R.
Hence $\rho_{21|\L}\rho_{12|\R}-\rho_{12|\L}\rho_{21|\R}$ in
(\ref{kati}) is (proportional to) the probability that the particle came
from L and leaves to R minus the probability of the reverse sequence of
events. 

Eqs.~(\ref{1}--\ref{4.0}) imply for the stationary values
\BEA
\label{7.0}
\J_\L=\frac{\sinh[\frac{\mu_\L-\mu_\R}{2}]}{\tau_\L\cosh[\frac{\Delta E-\mu_\R}{2}]
+\tau_\R\cosh[\frac{\Delta E-\mu_\L}{2}]},\\
\label{7.7}
\p_1=\left[1+ e^{-\Delta E+l+\frac{\mu_\L+\mu_\R}{2}} \right]^{-1}, \quad \p_2=1-\p_1,\\
l\equiv \ln\left[({\tau_\L+\tau_\R e^{\frac{\mu_\L-\mu_\R}{2}   } })\left/({ \tau_\R+\tau_\L
e^{\frac{\mu_\L-\mu_\R}{2}   } })\right]\right. ,
\label{7.77}
\EEA
where $\Delta E$ is defined in (\ref{oro}).

Now $\J_\L\geq 0$ for $\mu_L\geq\mu_\R$: the stationary current
is from the higher chemical potential to the lower one. 

Note from (\ref{7.7}) that provided that $l=0$, the stationary
probability $\p_1$ of $\F$ has a Gibbsian form with the temperature
$T=1$ and chemical potential $\frac{\mu_\L+\mu_\R}{2}$; $l=0$ is
achieved under $\mu_\L=\mu_\R$ (overall equilibrium) or
$\tau_\L=\tau_\R$. If the latter condition holds, 
$\frac{\mu_\L+\mu_\R}{2}$ is in between of the
chemical potentials of the reservoirs $\L$ and $\R$; hence $\F$ is
generally not in equilibrium with them, even when it has a Gibbsian
form. The fact of having a Gibbsian form|that physically means the
existence of a local equilibrium|will be seen below to have important
consequences.

\subsection{Maximization of the particle current}
\label{maxima}

We want to have the largest $\J_\L$ for given $\mu_\L$ and $\mu_\R$, because this means
the optimal functioning of the matter-transporting machine. We start with a general 
premise that the current $\J_\L$ should be finite. This implies from (\ref{7.0}) two 
natural constraints [${\cal T}$ is a constant, k=L,R]:
\BEA
\label{kor}
\tau_\k\geq {\cal T}, 
\EEA
i.e. time-scales cannot be too short. The largest $\J_\L$ for given
$\mu_\L$, $\mu_\R$ and (\ref{kor}) is obtained when we maximize
(\ref{7.0}) with all the three involved parameters $\tau_\L$, $\tau_\R$,
$\Delta E$ being independent under conditions (\ref{kor}):
\BEA
\label{star}
\J_\L^* \equiv  \max_{\tau_\k,\Delta E}\,[\J_\L]
= \frac{1}{{\cal T}}\sinh[\frac{\mu_\L-\mu_\R}{4}],
\label{maxi}
\EEA
which is reached for the optimal values
\BEA
\label{baxi}
\tau^*_\L=\tau^*_\R={\cal T}, \qquad \Delta E^*=\bar{\mu}\equiv (\mu_\L+\mu_\R)/2.
\label{mumu}
\EEA
Eq.~(\ref{star}) is got as follows. We maximize $\J_\L$ over $\tau_\L$ and $\tau_\R$
and obtain the first condition in (\ref{baxi}). Next we maximize over $\Delta E$.

Here are the implications of (\ref{star}--\ref{baxi}).

{\bf 1.} The optimal $\tau^*_\k$ are fixed by the
constraints. In contrast, $\Delta E^*$ depends on the environment
(reservoirs). The optimality condition $\tau_\L=\tau_\R$ implies $l=0$
from (\ref{7.77}). Hence the stationary probabilities $\p_i$ for $\F_i$
have the Gibbsian form with chemical potential (\ref{mumu}); see
(\ref{7.7}) and the discussion after (\ref{7.77}). This consequence of the current
maximization is one the main causes of our results below. 

{\bf 2.} The machine performing optimally in one environment will be
sub-optimal in another. Indeed, let the parameters be fixed at their
optimal values (\ref{baxi}), and the chemical potentials are slowly
changed as
\BEA
\mu_\L\to\mu'_\L,~~\, \mu_\R\to\mu'_\R,~~\, \mu'_\L>\mu'_\R, ~~\, \bm\to\bm'\equiv\bm+\delta,\,
\label{env}
\EEA
where $\delta=\frac{1}{2}(\mu'_\L - \mu_\L + \mu'_\R-\mu_\R)$.

The stationary current in the new situation (\ref{env}) is 
\BEA
\label{caro}
\J_\L
[\mu'_\L,\mu'_\R]={\J_\L^* [\mu'_\L,\mu'_\R]}/{\cosh[\,\frac{\delta}{2}\,]}, 
\EEA
where $\J_\L^* [\mu'_\L,\mu'_\R]$ is the optimal current in the new
environment; see (\ref{baxi}, \ref{7.0}).  For $|\delta|\gg 1$ we get
${\J_\L [\mu'_\L,\mu'_\R]}\ll{\J_\L^* [\mu'_\L,\mu'_\R]}$: the current
becomes suboptimal for any sizable environmental change. If 
the chemical potential difference is conserved,
$\mu'_\L-\mu'_\R\approx\mu_\L-\mu_\R$, the current will decrease from
its old value: ${\J_\L[\mu'_\L,\mu'_\R]}\ll{\J_\L^* [\mu_\L,\mu_\R]}$. 

{\bf 3.} The formal reason of this fragility is that the optimal $\Delta
E^*$ depends on the environment. Its physical reason is that $\F$ has
to perform equally well two complementary things: to bind and release.
To illustrate this point, recall that $N_2=1$, $N_1=0$ and
assume in (\ref{baxi}, \ref{6.0}):
\BEA
\label{mangoost}
\tau_\L=\tau_\R={\cal T}=1.
\EEA
Then the binding transition $\F_1\to \F_2$ is driven mainly by the L
reservoir, $\rho_{21|\L}>\rho_{21|\R}$, while the releasing transition
$\F_2\to \F_1$ is driven mainly by R: $\rho_{12|\R}>\rho_{12|\L}$ (this 
is why $\F$ transports particles from L to R). In
the optimal regime both $\rho_{21|\L}$ and $\rho_{12|\R}$ should be
possibly large.  Hence they are equal:
$\rho_{21|\L}=\rho_{12|\R}=e^{\frac{1}{4}(\mu_\L-\mu_\R)}$, i.e. $\F$
binds and releases the particle equally well.  Now after the
environmental change (\ref{env}):
$\rho_{21|\L}/\rho_{12|\R}=e^{{\delta}/{2}}$. Thus for $\delta>0$
($\delta<0$) the current is sub-optimal, since $\F$ binds the particle
better (worse) than releases it. 

\comment{This is also seen from (\ref{7.7}), where $\p_1=\p_2=\frac{1}{2}$ under conditions (\ref{baxi}).}

{\bf 4.} Our model relates to inhomogeneous
catalysis. Any catalysis facilitates the spontaneous transfer of
reacting molecules from a higher to a lower chemical potential
\cite{chem}. During an inhomogeneous catalysis the reactants are bound
strongly to an active center of the catalysing surface, so that the
reaction can proceed. But the reaction products should be weakly bound
to the center, so that they are easily released making the center ready
for a new reaction. This complementarity between binding and releasing
is essential for any good catalyst; e.g. silver and tungsten are both not
good catalysis for organic molecules: silver binds reactants too weakly,
while tungsten binds the products too strongly \cite{chem}. 

Another situation, where binding and releasing are simultaneously
important is the oxygen transport by mammal erythrocytes. They are
periodically removed from the blood, since they cease to perform well
one of these functions, e.g., they bind oxygen too strongly \cite{trincher}. 

{\bf 5.} Recall that in deriving (\ref{maxi}) we assumed that the
current $\J_\L$ can be maximized over independent parameters $\tau_\L$,
$\tau_\R$, $\Delta E$; see also {\bf 2}, where this
assumption was used implicitly. While this leads to the largest value of
$\J_\L$|in the sense that any relation between the three
parameters can only reduce the optimal value (\ref{maxi})|it is still
possible that $\tau_\k$ does depend on $\Delta Q_\k=\Delta E-\mu_\k$, 
e.g. in the activated transport \cite{weiss}
\BEA
\label{sta}
\tau_{\rm k}=\exp[v_{\rm k}+\frac{1}{2}|\Delta Q_\k|],
\EEA
where $v_{\rm k}$ is the barrier height. We now show that in the
linear regime $\mu_\L\approx\mu_\R$ we can recover the same conclusions
as above without assuming that $\tau_\k$ does not depend on $\Delta
Q_\k$. 

In the linear regime we put $\mu_\L=\mu_\R=\bm$
everywhere besides $\sinh[\frac{\mu_\L-\mu_\R}{2}]\approx\frac{\mu_\L-\mu_\R}{2}$ in
(\ref{7.0}):
\BEA
\label{7.1}
\J_\L=\frac{\mu_\L-\mu_\R }{2[\tau_\L(\Delta Q)
+\tau_\R(\Delta Q)]\cosh[\frac{\Delta E-\bm}{2}]},
\EEA
where $\tau_\L$ and $\tau_\R$ are functions of $\Delta Q=\Delta E-\bm$ \cite{weiss}.
They have global minima at $\Delta Q=0$ [see our discussion after
(\ref{66.00})]. In the linear regime the stationary probabilities of $\F$
are naturally Gibbsian, since $l=0$ in (\ref{7.7}, \ref{7.77}).

Maximizing (\ref{7.1}) over $\Delta E$ we get
$\J_\L^*=\frac{\mu_\L-\mu_\R}{2[\tau_\L(0) +\tau_\R(0)]}$ reached
for $\Delta E^*=\bar{\mu}$.  In the linear regime
these agree with (\ref{maxi}). Also, we reproduce the conclusions of
{\bf 1-4} by keeping the dependence of $\tau_\k$ on $\Delta Q$.  Note
that the changes in (\ref{env}) should respect the linear regime:
$\mu'_\L-\mu'_\R\approx\mu_\L-\mu_\R\approx 0$. 

{\bf 6.}
We aimed to show that in the linear regime our conclusions on the
current optimality and its fragility apply more generally.  We do not
assume the linear regime for the rest of this paper.  Below we set
$\tau_\L=\tau_\R={\cal T}$, because this maximizes $\J_\L$ with all other
parameters being fixed. For technical simplicity we from now on put ${\cal T}=1$
[cf. (\ref{mangoost})]. 

\comment{\begin{gather}
\p_1=\left[1+ e^{-\Delta E+\kappa+\frac{[\mu_\L+\mu_\R]\Delta N}{2}} \right]^{-1}, \quad \p_2=1-\p_1, \\
\kappa\equiv \ln\left[ \frac{1+e^{\frac{A_\L-A_\R+[\mu_\L-\mu_\R]\Delta N}{2}   } }{ e^{\frac{A_\L-A_\R}{2} }+
e^{\frac{[\mu_\L-\mu_\R]\Delta N}{2}   } }  \right],
\end{gather}}

\comment{
For definitness we assume from now on that $A_\k$ are constants, i.e.
they do not depend on $\Delta E$ or $\Delta N$. Qualitatively same
results are obtained for our other reasonable choices of $A_\k$, e.g.,
$A_\k=2V-[\Delta E-\mu_\k\Delta N]_+$, where $[x]_+=x$ for $x\geq 0$ and
$[x]_+=0$ otherwise. This choice of $A_\k$ in (\ref{6.0}) leads to activated
transitions $i\to j$ with the barrier $V$. }

\section{Structural degree of freedom} 
\label{structural}

\subsection{Master equation}

\begin{figure}
\includegraphics[width=7cm]{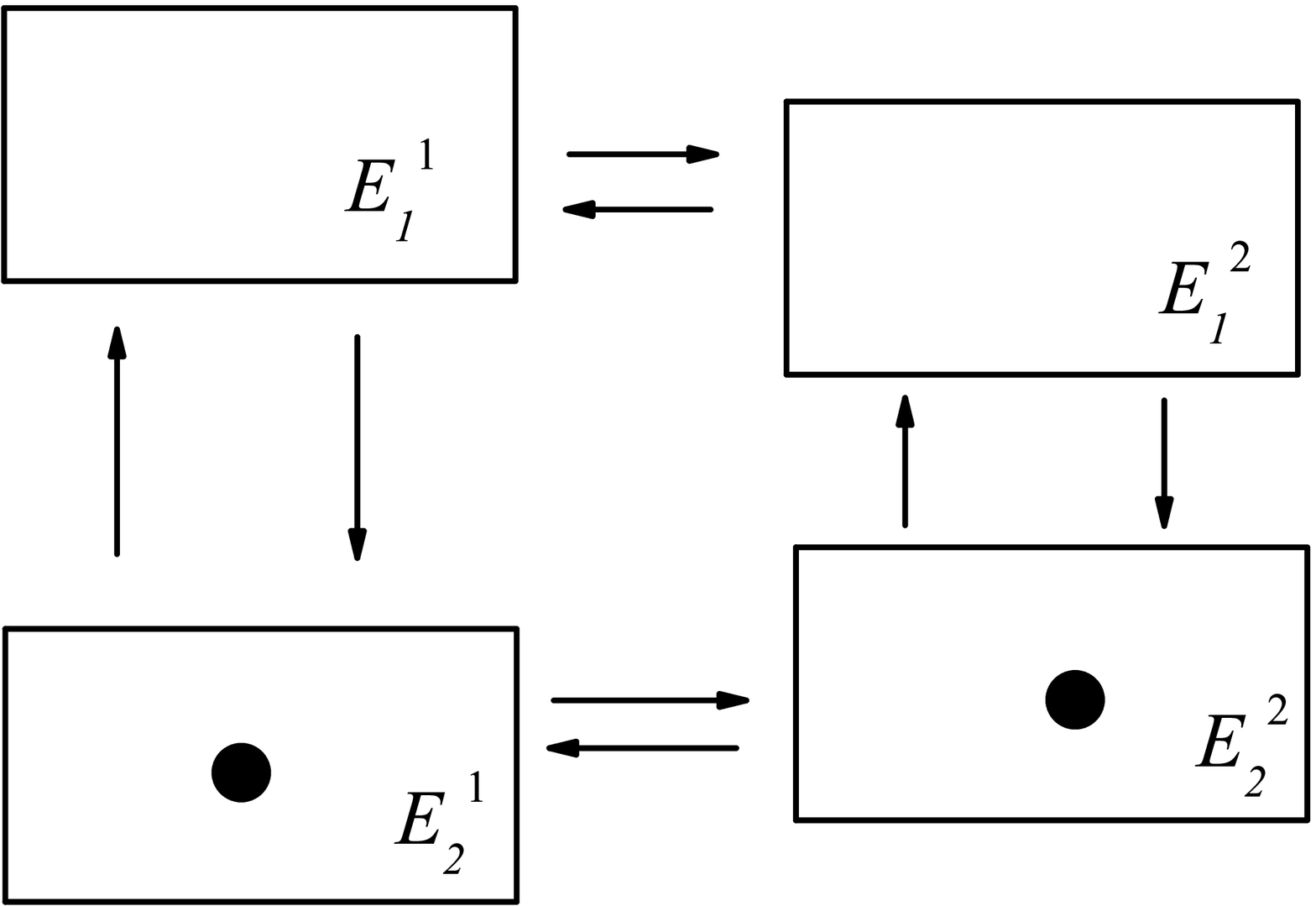}
\vspace{0.8cm}
\caption{ The joint states of functional degree $\F$ and the structural degree $\S$. For simplicity it is assumed
that both $\F$ and $\S$ can have two separate states ($\F_1,\F_2$ and $\S^1,\S^2$, respectively); recall
Fig.~\ref{fig1}. Hence there are
four states $\F_i\S^\alpha$ (of $\F+\S$) with energies $E_i^\alpha$. The transitions $\F_i\S^\alpha\to \F_i\S^\beta$,
$\F_i\S^\alpha\to \F_j\S^\alpha$ take place between states having one common index; see (\ref{4}). This is due to the
fact that the reservoirs that drive the transitions are independent from each other.
} 
\label{fig3}
\end{figure}

We discussed in {\bf 2} that the environmental changes
(\ref{env}) diminish the optimal current. If our machine is supposed to
work in such an uncertain environment, the only possibility to ensure
its optimal functioning is to assume that its structure $\S$ changes and
adjusts the energy difference to $\Delta E=\bm'$ after each
environmental change $\bm\to\bm'$; see (\ref{env}). Then the current is
maximal under each environment. To account for structural changes we
thus introduce a controller degree of freedom $\S$, with states
$\{\S^\alpha\}_{\alpha=1}^K$; see Fig.~\ref{fig3}. We shall demand that 
$\S$ is slower than $\F$ and that it does not couple {\it directly} to
the changing environment.

Let $p_i^{\alpha}$ and $E_i^{\alpha}$ be, respectively, the joint
probability and energy of $\S^\alpha\F_i$. Since $\S$ does not couple to
particle reservoirs L and R, each state $\S^\alpha\F_i$ carries the
number of particles $N_i$ that does not depend on $\alpha$; see
Fig.~\ref{fig3}. Hence the $\S-\F$ coupling goes only via the energies
$E_i^{\alpha}$. 

The dynamics of $\S$ is driven by a thermal bath at a temperature
$1/\theta$; see Fig.~\ref{fig2}. 
It will be seen below that the adaptation makes necessary for the 
baths of $\F$ and $\S$ to be different.
Hence we assume that the baths of $\S$ and $\F$ are
independent, and the general master equation $\dot{p}_i^{\alpha} =
{\sum}_{j\gamma}[R_{ij}^{\alpha\gamma}
p_j^{\gamma}-R_{ji}^{\gamma\alpha}p_i^{\alpha}]$ for $\S+\F$ reduces to
\BEA
\dot{p}_i^{\alpha} = {\sum}_{j}[\rho_{ij}^{\alpha}p_j^{\alpha}-\rho_{ji}^{\alpha}p_i^{\alpha}]+\epsilon 
{\sum}_{\gamma}[\omega^{\alpha\gamma}_{i}p_i^{\gamma}-\omega^{\gamma\alpha}_{i} p_i^{\alpha}],
\label{4}
\EEA
where $\omega^{\alpha\gamma}_{i}$ and $\rho_{ij}^{\alpha}=\sum_{\k=\L,\R}\rho_{ij|\k}^{\alpha}$ are the rates
of transitions $\F_i\S^\gamma\to\F_i\S^\alpha $ and
$\F_j\S^\alpha\to\F_i\S^\alpha$, respectively.
$\epsilon$ is the ratio between the
time-scales of $\F$ and $\S$. The detailed balance for
$\rho_{ij|\k}^{\alpha}$ reads [cf. (\ref{6.0})]
\BEA
\label{5.0}
\rho_{12|\k}^{\alpha}=e^{\frac{1}{2}(E^{\alpha}_2 -E^{\alpha}_1-\mu_{\k}   )}, ~~ \rho_{21|\k}^{\alpha}=1/\rho_{12|\k}^{\alpha},
\EEA
where in (\ref{5.0}) we used the settings (\ref{mangoost}); recall {\bf 6} in section \ref{maxima}. 

The detailed balance condition for $\omega^{\alpha\gamma}_{i}$ is written down by analogy to 
(\ref{6.0})
\BEA
\label{5}
\omega_i^{\alpha\gamma}=e^{\frac{1}{2}B^{\alpha\gamma}_i + \frac{\theta}{2}(E^\gamma_i -E^\alpha_i   )}, \quad 
B^{\alpha\gamma}_i=B^{\gamma\alpha}_i,
\EEA
where $B^{\alpha\gamma}_i$ relates to the inverse time-scale, and 
$\theta$ is the inverse temperature of the bath of $\S$ [recall (\ref{staro}) in this context].

\subsection{Time-scale separation}

It is generally understood that control processes in biology involve
time-scale separations between controlling and functional degrees of
freedom; see \cite{roj,gun} for reviews \footnote{The reasons for a
widespread applicability of the time-scale separation are summarized as
follows. {\it (i)} It allows to reduce the complexity of the overall
problem by separating (on fast times) the involved degrees of freedom
into statistical (functional) and mechanical (structural)
\cite{roj,gun}. In particular, this means that the stability of the fast
subsystem is determined under fixed values of the slow degrees of
fredom, a fact that we stress before (\ref{dur}). {\it (ii)} The evolution of slow
degrees of freedom is robust with respect to those parameters of the
fast subsystem that govern its dynamics, but do not show up explicitly
in the (conditional) stationary probabilities that determine the
effective transition rates of the slow subsystem \cite{roj}; see
(\ref{usa}) in this context.}. In line with this, we assume that $\S$ is
much slower than $\F$ for $\epsilon\to 0$. One introduces in (\ref{4})
the conditional probability $p_{i|\alpha}$ defined via
$p_{i|\alpha}p^\alpha=p_i^\alpha $, and notes that in
$\dot{p}_i^\alpha=\dot{p}^\alpha p_{i|\alpha}+p^\alpha
\dot{p}_{i|\alpha}$ we have $p^\alpha \dot{p}_{i|\alpha}={\cal O}(1)$
and $\dot{p}^\alpha p_{i|\alpha}={\cal O}(\epsilon)$ \cite{a+n}. Hence
$\dot{p}_{i|\alpha}$ and $\dot{p}^{\alpha}$ decouple:
\begin{gather}
\label{6.00}
\dot{p}_{i|\alpha} = {\sum}_{j\not= i}[\, \rho_{ij}^{\alpha}p_{j|\alpha}-\rho_{ji}^{\alpha}p_{i|\alpha}\,],\\
\label{6}
\dot{p}^{\alpha} = \epsilon{\sum}_{\gamma}[\Omega^{\alpha\gamma}p^{\gamma}-\Omega^{\gamma\alpha}p^{\alpha}], \,\,
\Omega^{\alpha\gamma} = {\sum}_i \omega^{\alpha\gamma}_i\,p_{i|\gamma}.
\end{gather}
Eq.~(\ref{6.00}) describes the evolution of $\F$ for short times when
$\S$ is fixed in its state $\S^\alpha$. Then $\F$ relaxes to its
conditional equilibrium$\p_{1|\alpha}=\frac{\rho_{12}^\alpha}{\rho_{12}^\alpha+\rho_{21}^\alpha}=[1+e^{-\Delta
E^\alpha+\bm}]^{-1}$, where $\Delta E^\alpha=E^\alpha_2-E^\alpha_1$; see
(\ref{stati}, \ref{7.7}).  Since this relaxation happens faster than
$\S$ changes, for describing the dynamics of $\S$ one can replace in
(\ref{6}):
\BEA
\label{usa}
\Omega^{\alpha\gamma}\to \bar{\Omega}^{\alpha\gamma} = {\sum}_i \omega^{\alpha\gamma}_i\,\bar{p}_{i|\gamma}. 
\EEA
Then (\ref{6}) becomes a Markov master equation for $\S$: the future of
$\S$ is determined by its own present state only. $\F$ influences $\S$
{\it indirectly} via the transition rates $\bar{\Omega}^{\alpha\gamma}$. 

Since most of the time $\F$ is in its stationary situations
corresponding to a fixed $\S^\alpha$, the functioning of the machine is
described by the current $\J^\alpha_\L$ in the states $\S^\alpha$ and
the stationary probability $\p^\alpha$ of $\S^\alpha$ found from
(\ref{6}, \ref{usa}). This is the main consequence of the time-scale
separation. 

Using (\ref{7.7}) with $l=0$ [see (\ref{6})] we get from (\ref{usa}, \ref{5}):
\begin{gather}
\frac{\bar{\Omega}^{\a\b}}{\bar{\Omega}^{\b\a}}=
e^{\theta(E_1^\b-E_1^\a)}\,\frac{1+e^{\bar{\mu} -\Delta E^\a }}{1+e^{\bar{\mu}-\Delta E^\b}}\,\,\times \nonumber\\
\frac{1+e^{b^{\a\b}+\frac{\theta}{2}(\Delta E^\b-\Delta E^\a)-\Delta E^\b+\bm   }  }
{1+e^{b^{\a\b}+\frac{\theta}{2}(\Delta E^\a-\Delta E^\b)-\Delta E^\a+\bm   }},
\label{katod}\\
\Delta E^\a \equiv E_2^\a-E_1^\a, \quad b^{\a\b}\equiv (B_2^{\a\b}-B_1^{\a\b})/2.
\label{katod2}
\end{gather}

At this stage we need to specify the transition rates
$\omega_i^{\alpha\gamma}$, because on one hand we want to have a
non-trivial structure of $\S$ (including the possibility of taking
$K\to\infty$), while on the other hand we want to get $\p^\a$ explicitly.
We thus choose to work with the birth-death model for $\S$
\cite{kampen}: the states $\{\S^\alpha\}_{\a=1}^K$ with energies
$\{E^\alpha_i\}_{\alpha=1}^K$ form a one-dimensional chain. In
(\ref{4}--\ref{6}) we allow transitions only between neighbouring
states: $\S^{\a}\leftrightarrows \S^{\a+1}$, $\a=1,\ldots,K-1$. Then the
stationary probability of $\S$ reads from (\ref{6}, \ref{usa})
\BEA
\label{dur}
\p^\alpha=\p^{1}{\prod}_{\gamma=1}^{\alpha-1} \left ( \bar{\Omega}^{\gamma+1,\,\gamma}/\bar{\Omega}^{\gamma,\,\gamma+1}\right), 
~~\alpha=2,\ldots,K,
\EEA
where $\p^{1}$ is determined from ${\sum}_{\alpha=1}^K \bar{p}^{\alpha}=1$.

\comment{&& \J^1_\L[\bar{\mu}_1] > \J^2_\L[\bar{\mu}_1], ~~~~ \p^1(\bar{\mu}_1)>\p^2(\bar{\mu}_1), \\
&& \J^2_\L[\bar{\mu}_2] > \J^1_\L[\bar{\mu}_2], ~~~~~~ \p^2(\bar{\mu}_2)>\p^1(\bar{\mu}_2).}

\section{Adaptation} 
\label{adaptation}

\subsection{Definition of adaptation}
\label{defiad}

The intuitive notion of adaptation is that this is a change in the
system compensating environmental effects. The general definition of
adaptation was attempted in literature several times and led to
interesting discussions; see \cite{formal,taivo} for reviews \footnote{It is sometimes said
that the fact of adaptation depends 
on the level of description; see \cite{taivo}.
This statement can be illustrated via a relaxator, a
system that relaxes to its final stationary state from a set of initial
states.  If the relaxator is perturbed to one of the initial states it
relaxes back to the stationary state. The perturbation is viewed
as an environmental change which is compensated when the relaxator goes
back to the stationary state. If one stays at the phenomenological
description of the relaxator, calling its relaxation process adaptation
amounts to trivialities. But the things are not anymore trivial
if we take into account that each relaxation is accompanied by a
change of some other quantity, which is hidden in the
phenomenological description, but within a deeper description
corresponds, e.g. to the energy of the reservoir that couples to the
relaxator and ensures its specific behavior. The behavior of
this quantity already deserves to be analyzed from the viewpoint of
adaptation.}.
But within
general definitions some important aspects of adaptation are left open,
the main of them is why (let alone how) the system is going to
compensate environmental effects \footnote{The ``how" question is also an important one.
In this context, one should distinguish robustness from adaptation. The first concept includes
stability mechanisms that do not lead to structural modifications in the system. These mechanisms are more like shielding the system from external perturbations.
The problem of combining robustness with efficiency was studied recently in
transport models of cell biophysics \cite{melk}.}.

In our situation the adaptation is a structural change needed for
compensating parametric environmental uncertainty which is detrimental
to the machine function: transporting particles with the maximal current
[see (\ref{caro}) and the discussion around]. 

Let the environment changes as in (\ref{env}),
and $\bm$ can assume any value from a set ${\cal M}$. It resides in each
of its states with a fixed $\bm$ for a sufficiently long time so that
$\F$ and $\S$ do have enough time to relax to the stationary
probabilities $\p^\alpha$ and $\p_{i|\alpha}$. We require that the
machine functions optimally, $\J_\L(\bm)=\J^*_\L$, in each environment:
the probability $\p^\a$ of the structural state with $\Delta
E^\alpha=\bm $ [see (\ref{baxi}, \ref{mangoost})] is maximized for each
$\bm\in {\cal M}$, i.e.  the probabilities of all other states are
suppressed in the sense discussed below. The choice between the optimal
structural states is done autonomously: after $\bar{\mu}$ changes from
one value to another (or goes back to its older value), $\F+\S$ relax to
a new stationary regime, where the structural state with the largest
current dominates. Recall that $\S$ does not feel the parameter
$\bar{\mu}$ directly, because it does not interact directly with the
particle reservoirs; it feels $\bar{\mu}$ only indirectly due to
interaction with $\F$. Hence the changes of $\S$ are driven by $\F$.

\subsection{Continuum limit}
\label{conti}

We assume that ${\cal M}$ is a finite real interval. There should be a
correspondence between $\{\S\}_{\alpha=1}^K$ and environmental states;
thus to have adaptation for this case we need to make $\alpha$ a
continuous variable $x$ and to take in (\ref{katod}--\ref{dur}) the
continuum limit: $K\to \infty$,
\BEA
\label{bo1}
E^{\alpha+1}_i-E^{\alpha}_i=E_i(x+\frac{1}{K})-E_i(x)=\frac{1}{K}
\partial_x E_i(x),\\
\label{bo2}
b^{\a,\a+1}=b(x,x+\frac{1}{K})=b(x)+\frac{1}{K}\partial_y b(x,y)|_{y=x}, 
\EEA
where $b(x)=b(x,x)$ and where
the continuous variable $x$ changes in an interval: $x\in [L_1,L_2]$.
Thus the stationary probability of $\S$ (which is now a probability density due to the continuum limit)
depends on three functions $E_i(x)$ and $b(x)$. 
These functions look arbitrary, but we show below that
they are fixed from the adaptation condition.

In the continuum limit the sums over $\gamma$ in (\ref{dur}) can be
replaced by integrals. As shown in Appendix \ref{pror}, the stationary probability of $\S$ then
reads from (\ref{dur}) [$A'(x)\equiv\partial_xA(x)$]
\begin{gather}
\label{gu}
\p(x) =\exp[-\Psi(x)]/Z, \\
\label{gugu}
\Psi'(x)=\theta f'(x)+(\theta-1)\phi(x)[f'(x)-E_1'(x)],\\
\label{gugugu}
 f(x)=-\ln[\,e^{-E_1(x)}+e^{-E_2(x)+\bm}\,],\\
\phi(x)=[\,e^{b(x)}-1\,]\,[\,e^{b(x)+\bm- E_2(x)+E_1(x)}+1\,]^{-1}, 
\label{dosh}
\end{gather}
where $Z$ in (\ref{gu}) is deduced from $\int_{L_1}^{L_2}\d x\,\p(x)=1$.
In (\ref{gugugu}), $f(x)$ is the free energy of $\F$ calculated at a
fixed $x$; see (\ref{7.7}) and \cite{a+n}. The second term in
(\ref{gugu}) is due to different temperatures of $\S$ and $\F$
($\theta\not =1$) and the dependence of the times-scales of $\S$ on $\F$
($b(x)\not =0$); see (\ref{5}, \ref{katod2}). 

For adaptation it is necessary that 
for any $\bm\in{\cal M}$, $\p(x)$ in (\ref{gu}) has a unique and sharp
maximum at $x=\x(\bm)$ with [cf. (\ref{baxi}, \ref{mangoost})]:
\BEA
\label{battre}
\Delta E(\x(\bm))\equiv
E_2(\x(\bm))-E_1(\x(\bm))=\bm.
\EEA
Then $\x(\bm)$ will be the most probable value of $x$.
Two conditions for $\x(\bm)$ to be a local maximum of
$\p(x)$ are $\Psi'(\x(\bm))=0$ and $\Psi''(\x(\bm))>0$.
Working out $\Psi'(\x(\bm))=0$ and using (\ref{battre}) we get
\BEA
\label{oparin}
\frac{e^{b(\x(\bm))} -1  }{e^{b(\x(\bm))} +1}=\frac{\theta}{1-\theta}\,\,\frac{ E'_1(\x(\bm))+E'_2(\x(\bm))  }{  E'_2(\x(\bm))-E'_1(\x(\bm))  }.
\EEA
This relation is supposed to hold for all $\bm\in {\cal M}$, and since
functions $E_{1,2}(x)$ and $b(x)$ do not depend on $\bm$|
because, in particular, $\S$ does not interact with the particle reservoir|
(\ref{oparin}) holds in a dense set of $x$. Once $E_1(x)$,
$E_2(x)$ and $b(x)$ are assumed to be smooth functions of $x$,
(\ref{oparin}) will hold for all $x\in [L_1,L_2]$ \, \footnote{The
precise statement is that a meromorphic (analytic up to isolated poles)
function can have only isolated zeros; otherwise it is equal to zero
\cite{iso}. If we assume that the difference between two sides of
(\ref{rashid}) is meromorphic, it is zero on a dense set due to
(\ref{oparin}) and hence is zero everywhere.}; it defines a functional
relation between $E_1(x)$, $E_2(x)$ and $b(x)$:
\BEA
\label{rashid}
\frac{e^{b(x)} -1  }{e^{b(x)} +1}=\frac{\theta}{1-\theta}\,\,\frac{ E'_1(x)+E'_2(x)  }{  E'_2(x)-E'_1(x)  }.
\EEA
We work now out $\Psi''(x)$ using (\ref{rashid}) and put there (\ref{battre}):
\begin{gather}
\Psi''(\x(\bm))=-\frac{[\Delta E'(\x(\bm))]^2}{4}\times\nonumber\\
\left[\,\theta+(1-\theta)\left[\frac{e^{b(\x(\bm))} -1  }{e^{b(\x(\bm))} +1}\right]^2 \,\right].
\label{zombi}
\end{gather}
We need $\theta<0$, i.e. a negative-temperature state of $\S$; otherwise
$\x(\bm)$ is a minimum of $p(\x(\bm))$, not maximum \footnote{Note that $\Psi(x)$ can have other minima and maxima. Our concern here
is the specific minima of $\Psi(x)$|and maxima of $p(x)$|that are given by (\ref{battre}). This condition is necessary for the adaptation. }. 
We now set $b(x)=0$, because
then any $\theta<0$ will suffice for adaptation; otherwise, $\theta$ has
to be smaller than certain negative value.  Eq.~(\ref{rashid}) under
$b(x)=0$ implies $E'_1(x)=-E'_2(x)$.  We shall see in section
\ref{flucto} that setting $b(x)=0$ appears to be optimal from the
viewpoint of reducing fluctuations of $x$ around its most probable
value. Recall from (\ref{katod2}, \ref{bo2}) the physical meaning of the
condition $b(x)=0$: the time-scale of $\S$ does not depend on the state
of $\F$. 

We conclude that the adaptation requires a negative temperature
$\theta<0$ for $\S$. This condition holds as well when the environmental
chemical potentials (and hence $\S$) assume a finite number of values;
see Appendix \ref{2state}. After bringing an explicit example of
adaptation, we continue with clarifying the generality of the $\theta<0$
condition (section \ref{chatchat}) and its physical meaning (section
\ref{nego}).

\subsection{An example}

Let us at this point stop the general reasoning and bring an explicit
example of above construction (\ref{gu}--\ref{zombi}): $E_i(x)=a_ix$ for
$i=1,2$ is linear function of $x$ with its slope $a_i$ dependent on $i$.
For a fixed $i$, $E_i(x)$ as a function of $x$ is a potential energy.
This potential energy is not confining, but it is not a problem, since
we anyhow restricted the allowed range of $x$: $x\in [L_1,L_2]$. Now
requiring $a_1=-a_2$ leads to $b(x)=0$; see (\ref{gu}--\ref{dosh},
\ref{rashid}). The stationary probability of $\S$ reads from
(\ref{gu}--\ref{dosh})
\BEA
\label{dodo}
\p(x)\propto (\cosh[a_2x-\frac{\bm}{2}])^{-|\theta|}.
\EEA
The most probable value of $x$ is $\frac{\bm}{2a_2}$, condition (\ref{battre}) is ensured, 
and the interval of allowed values of $\bm$ is
\BEA
\label{lunch}
{\cal M}=2a_2\times [L_1,L_2]. 
\EEA
Note that the range ${\cal M}$ of $\bar{\mu}$ over which the adaptation
occurs is finite and enters into the hardwire of $\S+\F$ as an apriori
knowledge on the environment. ${\cal M}$ is finite due to the fact that $\S$
interacts with a negative-temperature bath: if the range of $x$ is not
finite, such an interaction will lead to instability; see section \ref{nego}
and \cite{ramsey,butko,brunner}.

\subsection{Le Chatelier principle}
\label{chatchat}

Here we argue that the necessary condition $\theta<0$ for the adaptation
is more general that the above derivation may suggest. 

For $b(x)=0$ we get from (\ref{gu}--\ref{dosh}) that $\Psi(x)=\theta
f(x)$, where $f(x)$ is the free energy of $\F$ calculated for a fixed
slow variable $x$. The joint probability of $\F+\S$ is then
\BEA
\label{joker}
p_i(x)\propto e^{\theta f(x)}p_{i|x}, 
\EEA
where $p_{i|x}$ (as a function of $i$ for a fixed $x$) has a Gibbsian
form; see (\ref{7.7}, \ref{baxi}) and recall that in the optimal regime
$\tau_\L=\tau_\R$. Eq.~(\ref{joker}) for $p_i(x)$ implies the time-scale
separated two-temperature system, which admits a consistent
thermodynamic description despite the fact that the temperatures of $\S$
and $\F$ can be very different \cite{a+n}. For such a system, the fact
that the adaptation requires $\theta<0$ is deduced from the generalized
Le Chatelier principle: $\theta<0$ is necessary that perturbations of
the chemical potential {\it do not} amplify in time; see Appendix
\ref{chat}.

\subsection{Negative temperature states}
\label{nego}

Let us discuss in more detail the features of 
negative-temperature states. 

-- They maximize (not minimize) the average energy for a fixed entropy
\cite{ramsey,butko,brunner}. 

-- They thus store energy: a cyclic external field can extract work from
them \cite{ramsey,butko}. Such states are autonomous sources of work
\cite{brunner} \footnote{Note the difference: a system with a positive
temperature can still produce work (during cyclic action of external
field), if it is attached to a thermal bath at a different temperature;
cf. Footnote \ref{TT}.  In contrast, a negative-temperature system can
produce work autonomously, without external environment. Here the cyclic
condition means that the source of work couples and and then decouples
from the system.}.

-- To support $\theta<0$, $\S$ needs to have a bounded phase-space
\cite{ramsey,butko,brunner}, as already assumed \footnote{Hence the kinetic energy (which
is non-negative and can be arbitrary large) cannot be a part of those
degrees of freedom that support a negative temperature. }.

-- Negative temperatures are higher than positive ones, since heat flows
from lower inverse temperature to the larger one \cite{ramsey,butko,brunner}. This
implies that the energy stored in $\S$ due to $\theta<0$ will constantly
flow to the bath of $\F$ with the rate proportional to $\epsilon$ [see
(\ref{6}] tending to dissipate the stored energy. This current is
calculated in Appendix \ref{energy_disso}. Due to assumed $\epsilon\ll 1$, it
is smaller than the particle current $\J_\L$ in (\ref{7.0}). 

-- One should distinguish between population inversion (there are
at least two energy levels such that the higher energy level is more
populated than the lower one) and a negative-temperature state of a
many-level system, where in every pair of energy levels the higher
energy level is more populated. These two notions are equivalent for a
two level system, but generally they are different. A pair of energy
levels with population inversion suffices for (necessarily imperfect)
adaptation, if one can restrict $\S$ to those levels, i.e. neglect
transitions for higher energy levels; see Appendix \ref{2state}.
However, in that case the adaptation is necessarily imperfect, i.e.
there is no limit, where the probability of the undesired structural
states can be made arbitrary small. We focussed in section \ref{conti}
on the negative-temperature state of a many-level system, because it
does have a limit, where the adaptation can be made perfect; see section
\ref{flucto} below.

-- States with population inversion can be prepared by pumping the
system to higher energy levels; see \cite{butko} for an extensive review
of these methods.  This includes not only lasers and masers, but also
macroscopic magnetic moments, rotators, dipoles {\it etc} \cite{butko}.
It is also possible to prepare a population inversion via two weakly
coupled systems held at different positive temperatures \cite{brunner}
\footnote{\label{TT}Consider two very weakly coupled two-level systems
with energy level $(0,\epsilon_1>0)$ and $(0,\epsilon_2>0)$ held at
temperatures $T_1=\frac{1}{\beta_1}$ and $T_2=\frac{1}{\beta_2}$,
respectively. The joint system has four energy levels
$(0,\epsilon_1,\epsilon_2,\epsilon_1+\epsilon_2)$ with probabilities
$\propto (1,e^{-\beta_1\epsilon_1},
e^{-\beta_2\epsilon_2},e^{-\beta_1\epsilon_1-\beta_2\epsilon_2})$,
respectively.  The levels $\epsilon_1$ and $\epsilon_2$ of the joint
system display population inversionif$(\epsilon_1-\epsilon_2)(e^{-\beta_1\epsilon_1}-e^{-\beta_2\epsilon_2})>0$.
If in addition $\epsilon_1$ and $\epsilon_2$ are far from zero and close
to each other, the energy levels $\epsilon_1$ and $\epsilon_2$ of the
joint system can be regarded as decoupled from the rest of the spectrum
(for certain times).  }. 

-- Many-body states with a negative temperature are well known and were
experimentally realized for discrete and localized degrees of freedom
such as the spin of nuclei or atoms \cite{oja,butko}. One scenario of
their realization and observation is a microcanonic state of a many-spin
system in the regime, where the number of available states decreases
with increasing the energy \cite{oja,karen}. Negative-temperature states
were also experimentally realized for motional states of cold atoms
\cite{cold}, where the energy spectrum of the negative-temperature
carrying degrees of freedom is continuous, but bounded (similar to our
example in section \ref{conti}). 

-- There are complex materials held in metastable states that|without
being properly negative-temperature|behave like negative-temperature
states with respect to external variations \cite{ban}.

-- Living organisms are capable of being autonomous sources of work (the
work is done, e.g. for preventing their relaxation to equilibrium).
Hence they store energy at least via population inversion \footnote{In
living organisms the population inverted states is transferred from one
place to another via ATP \cite{blum}. It is an important and open
question whether also in ATP the energy is stored via population
inversion.  Alternatively, the ability ATP to produce work may be due to
different chemical potentials between ATP and its environment, which
means that ATP is not completely autonomous, it needs an environment to
produce work.  McClare argued that the relevant times of the ATP
work-delivery process are such that the coupling with the environment
can be neglected; hence, according to \cite{clare}, the energy is 
stored in ATP via population inversion.}. This fact is well understood in biological
thermodynamics \cite{bauer,clare,trincher_d,kauffman,voe} and was
employed in a definition of life \cite{kauffman}; see section \ref{reli} for more details.

\section{ Fluctuations}
\label{flucto}

We return to the stationary probability density $\p(x)$ of $\S$; see
(\ref{gu}).  Assume that the maximum $\x(\bm)$ of $\p(x)$ is unique.
There are local fluctuations around that maximum, since $\p(x)$ is not a
delta-function centered at $\x(\bm)$. Thus there are intrinsic
fluctuations of $x$, and thus of $\Delta E(x)$, even for an environment
with a fixed $\bm$. For $x\approx \x(\bm)$, (\ref{gu}) amounts to 
\BEA
\p(x)\propto \exp[-\frac{1}{2}|\Psi''(\x(\bm))|(\x(\bm)-x)^2],
\EEA
the standard deviation of $x$ is $\sigma_x=1/\sqrt{|\Psi''(\x)|}$. We
need to consider fluctuations of $\Delta E(x)$, since $\F$ feels $x$
only via $\Delta E(x)$. If $\sigma_x$ is small, 
the standard deviation of $\Delta E(x)$ around $\bm$ is
\BEA
\label{dolma}
\sigma=|\Delta E'(\x)|\sigma_{x}={2}/{\sqrt{|\theta|}}, 
\EEA
where we used (\ref{zombi}); recall that we set $b(x)=0$, this is also optimal 
from the viewpoint of reducing $\sigma$.

Now {\it the only} possibility of $\sigma\to 0$ is to take $\theta\to
-\infty$, i.e. a vanishing entropy of $\S$ due to a large amount of
stored energy. There are two general methods of fighting against
fluctuations: reducing the environmental noise as such (e.g. cooling the
environment), or putting the system under a strong confining potential.
The second method does not work in our situation, since it appears that
the strong potential cannot be adaptive. 

Thus, the adaptation is generally {\it imperfect}, because even at a
fixed environment (for a fixed $\bm$), $\Delta E$ will fluctuate on
characteristic times of $\S$.  The imperfect adaptation is useful only
if the environmental changes are wide enough; otherwise the same machine
with a fixed optimal structure tuned to the average environment will
have a larger average current. This point is worked out in Appendix
\ref{ap_flucto}. 

\section{ Alternative set-ups of adaptation} 
\label{alternative}

To get the proper perspective on the obtained results we briefly mention
certain extensions of the basic set-up. 

Above we required that the set ${\cal M}$ of environmental values of
$\bm$ is dense. The same limitations (negative temperature and
structural fluctuations) are obtained when the environment assumes a
finite number of values; see Appendix \ref{2state}. But here
fluctuations are finite even for $\theta\to-\infty$. 

Another set-up is to allow $\S$ to interact directly with the uncertain
environment [see Appendix \ref{no_cost}]. This is done naturally
assuming that different states $\F_i\S^\a$ have different particle
numbers $N^\a_i$. The above limitations are then absent: even the
two-state $\S$ (in face of a binary environment) can ensure perfect
adaptation in the isothermal situation $\theta=1$. In this set-up $\S$
becomes an {\it external sensor} for the machine. (This is sometimes
called feed-forward scenario of adaptation in contrast to the previous
scenario that works via feedback to $\F$: ${\rm\bf
environment}\to\F\to\S\to\F$.) Hence no stored energy is needed and no
fluctuations at the constant environments are necessarily present.
(Fluctuations may still be present due to various non-idealities.) The
drawback of this set-up is that it demands a specific
structure-environment interaction that has to designed anew for every
new environment.  Put differently, the above thermodynamic costs are
necessary for ensuring the autonomous character of the adaptation. 

Yet a different set-up is to relax the optimality condition demanding
instead the stabilization (constancy) of the current on the face of
environmental changes at some sub-optimal value; the above limitations
are then also weakened. The conceptual problem with relaxing the
optimality condition is that then the very adaptation may easily become
useless: who wants to keep a machine that does not function well in any
environment? It is clear however that the practical examples of
adaptation will be rather of this, non-optimal type, and an important
chapter of the future research on the physics of adaptation should
perhaps focus on understanding the trade-offs between optimality and
thermodynamic costs.

\section{Discussion}
\label{disco}

\subsection{Summary} 
\label{summary}

Thus our adaptive machine consists of two parts $\F$ and $\S$
(functional and structural degrees of freedom, respectively) and works
as follows; see Figs.~\ref{fig2}, \ref{fig1}, \ref{fig3}. If the
external chemical potentials are constant in time|and provided that the
fluctuations of $\S$ around its most probable value can be neglected|the
functional degree of freedom $\F$ transports particles with the maximal
current (speed) from the higher chemical potential to the lower one. 

Let now the chemical potentials change. After they settle in new values,
$\F$ will transport particles with much smaller current (speed) than it
is allowed under new values of the chemical potentials 
[see section \ref{functional}, {\bf 2}]. $\F$ will then
act on $\S$, and after a much longer time (much longer because $\S$ is
much slower than $\F$) $\S$ will feedback on $\F$ making the current
again maximal under the new values of the chemical potentials. For this
adaptation process to happen it is necessary that the structural degree
of freedom $\S$ has a sizable amount of stored energy, i.e. its
temperature is negative $\theta<0$. A negative $\theta$ is needed for
ensuring that perturbations induced by changing chemical potentials are
not amplified in time.  If, simultaneously, $|\theta|$ is sufficiently
large, the fluctuations of $\S$ around its most probable value are
negligible. 

However, because $\theta$ is necessarily finite, there are intrinsic
fluctuations of $\S$ that sometimes force the current to deviate from
its maximal value even under a constant environment. The origin of these
fluctuations inherently relates to the adaptation: it is impossible to
prevent fluctuations by confining $\S$ via an external potential, since
the latter will spoil the adaptation. 

On very long times|which are conceivable, but not directly seen on this
model|the second law will force $\theta$ to increase due to heat
exchange with the baths of $\F$ \cite{ramsey}. Hence $\S$ will make more
errors (fluctuate stronger) even in the constant environment; see
(\ref{dolma}). For even a larger $\theta$ having a variable structure
$\S$ may be useless: fluctuations of $\S$ are then so strong that in the
sense of the average current it is better to have a constant
(non-variable) structure [see Appendix \ref{ap_flucto}]. Thus,
adaptation is not always useful. 

\subsection{Relations with thermodynamic theory of aging}
\label{reli}

The functioning of the adaptive machine resembles certain features of
living organisms, in particular, the process of aging (senescence) that
is generally defined as progressive loss of stability of an organism
that increases the probability of its failure and that arises out of the
normal functioning of the organism \cite{agingreview}. Aging is a
complex phenomenon with many inter-related mechanisms at play; various
theories of aging emphasize different mechanisms \cite{agingreview}. We
shall focus on analogies between the functioning of the adaptive machine
and the thermodynamic theory of aging, whose different aspects were
uncovered in \cite{bauer,agingreview,voe}. The theory is one under
development, but it is already recognized in gerontology
\cite{agingreview,voe}. Since it is not well-known in statistical physics, we
shall briefly review the main postulates of this theory and then outline
their similarities with the adaptive machine studied above. 

\subsubsection{Postulates of the thermodynamic theory of aging}

{\bf 1.} Living organisms are in a non-equilibrium state that is
characterized by a certain amount of stored energy (i.e energy related
to population inversion) \cite{bauer,voe}; see section \ref{nego}. This
stored energy is needed for performing various function, e.g. the
organism needs it for preventing (doing work against) its own
equilibration \cite{bauer}.

{\bf 2.} For simplicity we shall visualize this stored energy as a
negative-temperature reservoir at temperature $-|\theta|$ (though in
reality it will most likely has more complex forms not described by a
single negative temperature). The reason why the non-equilibrium state
is characterized by a negative temperature (or population inversion, but
not just an excess free energy with respect of a given thermal
environment) was already explained in section \ref{nego}: the organism
should be capable to perform work autonomously. 

{\bf 3.} The stored energy is inborn, because the organism cannot
obtain it {\it only} from digesting (non-equilibrium) food \cite{bauer,voe}.
Indeed, digestion is a complex process that itself needs investment of
work \cite{bauer,voe}. It is assumed that some amount of stored energy is contained
already in the seed (for plants) or in the zygote (for mammals)
\cite{bauer,voe}.  \footnote{One of the most cited points on
non-equilibrium character of living organisms belongs to Schr\"odinger
who opined that organisms feed on negative entropy \cite{what}. This
point is questionable for two reasons; first, it does not recognize the
inborn character of the non-equilibrium (stored energy), second it does
not take into account that food contains mostly energy, not negative entropy
\cite{corning}.}

{\bf 4.} During the life of the organism, $|\theta|$ monotonically
decreases, due to various functions performed by the organism (and due
to inevitable coupling of the negative-temperature bath to
positive-temperature baths) \cite{bauer,voe}. Aging refers to state,
where $|\theta|$ is so small that the organism cannot perform adequately
its functions.  In particular, controlling systems of the organism get
progressively less stable and less efficient in responding to signals
from functional degrees of freedom. This point is emphasized in the
ontogenetic theory of aging \cite{dilman}. 

{\bf 5.} The free energy provided by food does not increase $|\theta|$,
it is only used for preventing its fast decrease and for extending the
size of the negative-temperature bath, i.e. the amount of stored energy
can increase, with its temperature $\theta<0$ still decreasing by its
absolute value \cite{bauer,voe}. The situation where the amount of the
stored energy increases corresponds to the growth of the organism
\cite{bauer,voe}. Hence a mature organism (the one that does not grow
anymore) has a limited amount of resource available for its functional
tasks; this point is fundamental for certain other theories of aging,
notably for the {\it disposable soma theory} \cite{agingreview}. 

\subsubsection{Relations with the adaptive machine}

The above points {\bf 1} is basic for the adaptive machine: the stored
energy in the structural degree of freedom ${\bf S}$ is needed for the
adaptive functioning of the machine (stability of the maximal current
under environmental changes). Also, the point {\bf 2} is there, since it
is only the simplest models of $\S$ that require a single and
well-defined negative temperature. 

The gradual decay of $|\theta|$ and the ensuing instability for the
adaptive machine|to an extent that having a complex, adaptive structure
is detrimental|resembles the aging process; see our discussion in
section \ref{summary} and points {\bf 3} and {\bf 4} above. Note as well
the relation with the ontogenetic theory of aging that stresses
progressive losses in controlling systems of the organism. 

For the adaptive machine the external non-equilibrium
($\mu_\L\not=\mu_\R$) environment cannot be used to support the
negative-temperature state of $\S$, we needed to assume a separate
negative-temperature thermal bath for $\S$ (points {\bf 3} and {\bf 4}
above). In our situation, the size of the bath of $\S$ is large but
fixed, which correponds to a mature (not growing) organism; see {\bf 5}
above. 

Recall that above analogies came out from combining a statistical
thermodynamics model with the stability of the maximal current under
environmental changes. Note that though the above conditions for
adaptation are obtained for a particular model, we argued that they hold
more generally, and are based on the quasi-equilibrium transport and the
Le Chatelier principle. 

\comment{
{\it Conclusion.|} We aimed to understand the minimal model of
self-adaptive machine that transports particles from one reservoir to
another. The model fits well to the phenomenology of heterogeneous
catalysis. The machine is supposed to function with the maximal current
under uncertain environment. This is achieved via structure-adaptation
and is possible under two limitations. First, some energy should be
stored in the structure. Second, the machine must sometimes deviate from
its maximal power even for a fixed environment.  Though the limitations
are obtained for a particular model, we argued that they hold more
generally, and are based on the quasi-equilibrium transport and the Le
Chatelier principle. }

\acknowledgements

We are grateful to Aram Galstyan for discussions. 
This work is supported by the Region des Pays 
de la Loire under the grant 2010-11967.

\comment{Open problems:

-- Which references to put in the structure-function relation for enzymes? The issue is controversial. First this is now new orthodoxy, 
while you have put here pretty "grim" references. Second once it became new orthodoxy, there appeared an opposition to it, a quite staunch
one. And what became an orthodoxy, cannot be completely right either. So I should decide what to do here. 

-- Do I want to mention the non-optimal regulation aspect or no. 

-- Is it not an abuse of "chemical terminology" to call "catalyst" a monomolecular reaction? Everybody knows that heterogeneous catalysis
and fermentative catalysis include second-order steps. So would not it simply lead to a confusion?

}

\appendix

\section{Derivation of Eqs.~(\ref{gu}--\ref{dosh})}
\label{pror}

Before applying (\ref{katod}) to (\ref{dur}) we change in (\ref{katod}) $\alpha\to\gamma+1$ and $\beta\to\gamma$. Next, applying
(\ref{bo1}, \ref{bo2}) we get for the first factor in (\ref{katod}) within the first order of the small parameter $\frac{1}{K}$:
\BEA
\label{aa1}
e^{\theta(E_1^\gamma-E_1^{\gamma+1})}\simeq e^{-\frac{\theta}{K}E'_1(x)}, 
\EEA
where $E_1(x)=E_1^{\gamma}$ and $E_1'(x)=\partial_x E_1(x)$.
Likewise, we obtain for other factors in (\ref{katod})
\BEA
\label{aa2}
&&\frac{1+e^{\bar{\mu} -\Delta E^{\gamma+1} }}{1+e^{\bar{\mu}-\Delta E^{\gamma}}}\simeq 
\exp\left[
-\frac{\Delta E'(x)}{K}\,\frac{e^{\bar{\mu}-\Delta E(x)} }{1+e^{\bar{\mu}-\Delta E(x)}}
\right],~~~~~\\
&&\frac{1+e^{b^{\gamma+1,\gamma}+\frac{\theta}{2}(\Delta E^{\gamma}-\Delta E^{\gamma+1})-\Delta E^{\gamma}+\bm   }  }
{1+e^{b^{\gamma+1,\gamma}+\frac{\theta}{2}(\Delta E^{\gamma+1}-\Delta E^{\gamma})-\Delta E^{\gamma+1}+\bm   }}\nonumber\\
&&\simeq 
\exp\left[
-\frac{(\theta-1)\Delta E'(x)}{K}\,\frac{e^{b(x)+\bar{\mu}-\Delta E(x)} }{1+e^{b(x)+\bar{\mu}-\Delta E(x)}}
\right],
\label{aa3}
\EEA
where $\Delta E=E_2(x)-E_1(x)$ and $b(x)=b^{\gamma+1,\gamma}+{\cal O}(\frac{1}{K})$; see (\ref{bo1}, \ref{bo2}).
Combining (\ref{aa1}--\ref{aa3}) into (\ref{dur}) and changing there $\frac{1}{K}\sum_{\gamma}\to \int\d x$ we get
(\ref{gu}--\ref{dosh}).

\section{Le Chatelier principle}
\label{chat} 

Here we shall derive the Le Chatelier principle for perturbations of the
chemical potential. In contrast to known derivations of the principle
that are purely thermodynamical, the present derivation stays within
statistical mechanics. We want to relate the principle to conditions
required for adaptation. 

Though the Le Chatelier principle is widely known and frequently applied
outside of physics (see \cite{aram} for further references on such
inter-disciplinary applications), already its thermodynamic derivation
contains several subtle points; see \cite{gilmore} for a review.  The
statistical mechanic derivation of the Le Chatelier principle for
perturbations of an intensive variable (chemical potential) is to our
knowledge presented for the first time. For perturbations of extensive
variables, the principle was recently discussed from the viewpoint of
kinetics (non-equilibrium statistical mechanics) \cite{aram}. 

We shall work within a set-up close to that in the main text.  There are
two interacting systems $\S$ and $\F$.  Now $\S$ is described by
coordinate (continuous variable) $x$, while $\F$ can be in discrete
states $i=1,...,n$. Each such state has energy $E_i(x)$ and carries
$N_i$ particles; recall our discussion in section \ref{defo}. The
coupling between $\S$ and $\F$ is due to dependence of $E_i(x)$ on $x$. 

$\F$ couples with a reservoir at temperature $1$ and chemical potential
$\bar{\mu}$.  $\S$ couples with an energy reservoir at inverse
temperature $\theta$. $\S$ is much slower than $\F$. We assume that the
conditions for the two-temperature adiabatic quasi-equilibrium hold|see
the discussion after (\ref{zombi}) and \cite{a+n}|which means that the
stationary probability $\p(x)$ of $\S$ and the conditional stationary
probability $\p_{i|x}$ of $\F$ read
\begin{gather}
\label{sup1}
\p_{i|x}=\frac{e^{-E_i(x)+\bar{\mu} N_i}}{Z(x)}, ~~ Z(x)={\sum}_{i} e^{-E_i(x)+\bar{\mu} N_i}, \\
\p(x)\propto e^{\theta \ln Z(x)}.
\label{supo}
\end{gather}
Thus, the conditional stationary probability of $\F$ has the Gibbsian
form with the chemical potential $\bar{\mu}$ and temperature $1$, while
the stationary probability $\p(x)$ of $\S$ is given by the Gibbs
distribution (at inverse temperature $\theta$) with the free energy
$-\ln Z(x)$ of $\F$ calculated at a fixed value of $x$. 

Recall that having two time-scales (slow and fast) is the basic premise
of the Le Chatelier principle that shows up in all its formulations; see
\cite{gilmore,aram}. If the two temperatures are equal, $\theta=1$,
(\ref{supo}) reverts to the usual Gibbs density.  Denote:
\BEA
\label{sup3}
\langle N \rangle_x\equiv {\sum}_{i}\p_{i|x} N_i, \\
\overline{\langle N \rangle_x}\equiv \int\d x \p(x)\langle N \rangle_x.
\label{sup303}
\EEA
Since $\S$ is slow and
$\F$ is fast, $\langle N \rangle_x$ characterizes the average particle number in $\F$ for
intermediate times, when the state $x$ of $\S$ is fixed. $\overline{\langle N \rangle_x}$
is the average of $\langle N \rangle_x$ over all states of $\S$; it naturally characterizes
the average particle number in $\F$ for long times, where fluctuations of $\S$ are essential.

Eqs.~(\ref{sup1}--\ref{sup3}) imply
\BEA
\label{gosh1}
\partial_{\bm} \overline{\, \langle N \rangle_x\,}- \overline{\,\,\partial_{\bm}\langle N \rangle_x\,\,}
=\theta\,\left[\,\overline{\,\langle N \rangle_x^2\,}-\overline{\,\langle N \rangle_x\,}^{\,2}\,\right].
\EEA
For $\theta>0$, (\ref{gosh1}) is non-negative. This is statement of the
Le Chatelier principle for perturbations of intensive variables
(chemical potential): they are {\it amplified} in time \cite{gilmore}.
Indeed, $\overline{\,\,\partial_{\bm}\langle N \rangle_x\,\,}$ in
(\ref{gosh1}) is the response of $\S+\F$ to a perturbation of
$\bar{\mu}$ that is much faster than $\S$, but much slower than $\F$.
$\partial_{\bm} \overline{\, \langle N \rangle_x\,}$ is the response of
$\S+\F$ to the same perturbation, but on much longer times so that
$\S+\F$ has thermalized at the perturbed value. Recall 
that for perturbations of extensive variables the statement of the Le Chatelier principle is just the opposite:
perturbations are suppressed (not amplified) in time \cite{gilmore,aram}. 

Now recall the set-up of section \ref{conti} and
assume that $\p(x)\approx \delta(x-\x(\bm))$ is a well localized around 
its average that is a necessary condition of perfect adaptation.
We now obtain for $\theta\geq 0$
\BEA
(\ref{gosh1})=\int\d x \frac{\partial \p(x)}{\partial\bm}\langle N \rangle_x
=\partial_{\bm}\x(\bm)\,\, \partial_x \langle N \rangle_x|_{x=\x(\bm)} \nonumber\\
=-
\partial_{\bm}\x(\bm)\,\,\Delta E'(\x(\bm))\,\,\partial_{\bm}\langle N \rangle_{\x}\geq 0, 
\EEA
where took into account that $\langle N \rangle_{x}$ depends on $x$ only
through $\Delta E(x)-\bm$.  Using $\partial_{\bar{\mu}}\langle
N\rangle_x\geq 0$ that also follows from the Gibbsian property
(\ref{sup1}) we get $\partial_{\bm}\x(\bm)\,\Delta E'(\x(\bm))\leq 0$
contradicting to the condition $\Delta E(\x)=\bm$ of adaptation; see (\ref{battre}).

\section{Quantifying imperfect adaptation}
\label{ap_flucto}

Section \ref{flucto} shows that the fact of structural fluctuations is an 
unavoidable consequence of adaptation.
We expect that such imperfect adaptation is not useful when the variance 
of environmental changes is small. To quantify this aspect, we
need to specify the statistics of environmental changes.
For simplicity we assume that in (\ref{env})
\BEA
\label{supsup}
\mu_\L-\mu_\R=\mu'_\L-\mu'_\R. 
\EEA
Hence $\bm=\frac{1}{2}(\mu_\L+\mu_\R)$ with probability $\P(\bm)$ is the only changing
environmental parameter. 

Let us now consider the current (\ref{7.0}) that is partially optimized:
we take in (\ref{7.0}) $\tau_\L=\tau_\R={\cal T}$, because this
maximizes $\J_\L$ with all other parameters being fixed; see
(\ref{kor}). For technical simplicity we fix the parameters as in
(\ref{mangoost}), i.e. we take ${\cal T}=1$. The resulting expression
for $\J_\L$ reads
\BEA
\label{sup_supo}
\J_\L[\Delta
E(x)-\bm]=\frac{\sinh[(\mu_\L-\mu_\R)/4]} {\cosh[(\Delta E(x)-\bm)/2]}.
\EEA 
We now average (\ref{sup_supo}) over over structural and environmental noise:
\BEA
J_1\equiv \int\d x\,\d\bm\, \J_\L[\Delta E(x)-\bm]\p(x)\P(\bm),
\EEA
where $\p(x)$ is the stationary state of $\S$; see section V. 

This is to be compared with the situation when the structure is fixed: first
$\Delta E$ is fit to its optimal value for one environment, e.g., the one
at $\langle\bm\rangle=\int\d\bm\,\bm \P(\bm)$, thus $\Delta
E=\langle\bm\rangle$ [recall (\ref{mangoost})]. Then the average over the environments is taken:
\BEA
J_2\equiv\int\d\bm\, \J_\L[\langle\bm\rangle-\bm]\P(\bm). 
\EEA
We expect that $J_2>J_1$ ($J_1>J_2$) whenever the width of $\P$ is sufficiently
smaller (larger) than that of $\p$; the adaptation is then useless
(useful). 

Let us now consider the example (\ref{dodo}): $E_i=a_ix$ for $i=1,2$ and
$x\in [L_1,L_2]$, where $a_1=-a_2$, $b=0$ and $\p(x)\propto
(\cosh[a_2x-\frac{\bm}{2}])^{-|\theta|}$. We assume that $\P(\bm)$ is
Gaussian with the standard deviation $d$ and mean $\langle\bm\rangle$.
We get
\BEA
\frac{J_1}{J_2}=\frac{\int\d x\, (\cosh[x])^{-|\theta|-1} }{\int\d x\, (\cosh[x])^{-|\theta|}}\,\,\frac{\sqrt{2\pi d}}
{\int\d x\, \exp(-\frac{x^2}{2d}-\ln\cosh[\frac{x}{2}])}.\nonumber
\EEA
Now $J_1/J_2$ is larger than one|and thus the adaptation is useful
for this example|if, e.g. $d=1$ and $|\theta|>4.811$. For $d\to\infty$,
$J_1/J_2\propto\sqrt{d}$ tends to infinity: the adaptation is
always useful for sufficiently large environmental uncertainty.

\section{Two-state environment}
\label{2state}

\subsection{Adaptation} 

Let the environment can be in two states $\E_{1}$ and $\E_{2}$, 
where $\bm=(\mu_\L+\mu_\R)/2$ assumes two different values
$\bm_1$ and $\bm_{2}$. These values are known, but it is not known
which one will be realized (cf. section \ref{defiad}). 
Then the minimal (necessary for adaptation) number of states for 
the structural degree of freedom $\S$ is also two, and
the structural and environmental states match as
\BEA
\E_\a\leftrightarrow \S^\a, ~~~E^\a_2-E^\a_1=\bm_\a, ~~~\a=1,2,
\EEA
where the second condition means that that each 
state $\S^\a$ provides the optimal value for $E_2-E_1$ [cf. (\ref{battre}, \ref{baxi}, \ref{mangoost})].
For the stationary probabilities of $\S$ we get [see (\ref{dur}) for $K=2$]
\BEA
\label{sup7}
\p^2=\p^{1}\bar{\Omega}^{21}/\bar{\Omega}^{12}, 
~~\p^1+\p^2=1.
\EEA
Using (\ref{sup7}) and (\ref{katod}) we obtain
\BEA
\label{nazar1}
\frac{\p^2(\bm_1)}{\p^1(\bm_1)}=e^{\theta(E_1^1-E^2_1)}\,\,\frac{1+e^m}{2}\,\,\frac{1+e^{b+\frac{\theta m}{2}}}{1+e^{b+(1-\frac{\theta}{2}) m}},\\
\frac{\p^1(\bm_2)}{\p^2(\bm_2)}=e^{-\theta(E_1^1-E^2_1)}\,\,\frac{1+e^{-m}}{2}\,\,\frac{1+e^{b-\frac{\theta m}{2}}}{1+e^{b-(1-\frac{\theta}{2}) m}},
\label{nazar2}
\EEA
where we defined [cf. (\ref{5}, \ref{katod2})] 
\BEA
\label{sup_polis}
m\equiv \bar{\mu}_1-\bar{\mu}_{2}, ~~~ b\equiv (B^{21}_2-B^{21}_1)/2.
\EEA
Eqs.~(\ref{nazar1}, \ref{nazar2}) imply
\begin{gather}
\frac{\p^1(\bar{\mu}_{2})}{1-\p^1(\bar{\mu}_{2})}\,\,\frac{\p^2(\bar{\mu}_{1})}{1-\p^2(\bar{\mu}_{1})}\nonumber\\
=
\frac{1+\cosh[m]}{2}\,\frac{\cosh[b]+\cosh[\frac{\theta m}{2}])}{\cosh[b]+\cosh[\frac{\theta m}{2}-m]   },
\label{dardo4}
\end{gather}

For the {\it perfect} adaptation it is necessary to have both
$\p^1(\bar{\mu}_{2})$ and $\p^2(\bar{\mu}_{1})$ going to zero (cf. section \ref{defiad}). 
Here $\p^1(\bar{\mu}_{2})$ and $\p^2(\bar{\mu}_{1})$ are the 
error probabilities, they characterize deviations from the optimal behavior
due to structural noise. If they both go to zero, then for
$\E=\E_1$ ($\E=\E_2$) one can neglect transitions to $\S^2$ ($\S^1$), at
least in the stationary regime, and the machine will function optimally for each
environment; see (\ref{battre}). 

The perfect adaptation is impossible:
the minimum of (\ref{dardo4}) equals to $1/4$ and it is reached for
\BEA
\label{go}
\theta<0,~~~~ b=0, ~~~~
|m|\gg 1, ~~~~ |\theta| |m|\gg 1.
\EEA
Indeed, $\theta<0$ is necessary for $(\ref{dardo4})<1$; the remaining 
conditions in (\ref{go}) is straightforward to obtain and interpret:
$b=0$ means that $\F$ does not alter the time-scales of
$\S$, while $|m|\gg 1$ means that the
environmental values are sufficiently different from each other. 
In this context recall our discussion on $b(x)=0$ after (\ref{zombi}).

The imperfect adaptation can be defined via
\BEA
\p^1(\bar{\mu}_{2})<\p^2(\bar{\mu}_{2}), ~~
\p^2(\bar{\mu}_{1})<\p^1(\bar{\mu}_{1}),
\label{dobro}
\EEA
i.e. the probabilities to get into ``wrong" structural states are
smaller than the probabilities to be in the ``right" states.
Eq.~(\ref{dobro}) means that the error probabilities
$\p^1(\bar{\mu}_{2})$ and $\p^2(\bar{\mu}_{1})$ are both smaller than
$1/2$. For this it is necessary that $\theta<0$; otherwise
(\ref{dardo4}) is always larger than $1$. Hence the imperfect adaptation
demands negative temperatures; cf. section \ref{conti}. 

For illustrating the imperfect adaptation let us assume that {\it i)}
$\mu_\L-\mu_\R$ stays constant [cf. (\ref{supsup})]; {\it ii)} $|m|\gg
1$; {\it iii)} both states $\E_1$ and $\E_{2}$ have equal probabilities
$1/2$. 

Now if the environment is in $\E_{\a}$ ($\a=1,2$), $\S$ is in its states
$\S^1$ and $\S^2$ with probabilities $\p^1(\bar{\mu}_{\a})$ and
$\p^2(\bar{\mu}_{\a})$, respectively.  If $\S$ is in $\S^\a$, the
particle current is given by $\J_L^*={\sinh[(\mu_\L-\mu_\R)/4]}$, see
(\ref{sup_supo}). If $\S$ is not in $\S^\a$, the particle current is
$\J_L^*/\cosh[m/2]$; cf. (\ref{caro}).  This current is negligible due
to the above condition {\it ii)}. Hence the current averaged over
environmental and structural fluctuations reads
\BEA
\label{ish}
\frac{1}{2}[\p^1(\bar{\mu}_{1})+\p^2(\bar{\mu}_{2})]\J_L^*.
\EEA
We compare (\ref{ish}) with a constant $\S=\S^1$ that is fit to the optimal
value $E^1_2-E^1_1=E^2_2-E^2_1=\bm_1$
of one environmental state.
In that situation the average
particle current is $\frac{1}{2}\J_L^*$. Requiring that
$\frac{1}{2}\J_L^*$ is smaller than (\ref{ish}) we get
$\p^1(\bar{\mu}_{1})+\p^2(\bar{\mu}_{2})>1$, which is ensured by the
definition (\ref{dobro}) of the imperfect adaptation. 

Concrete conditions for imperfect adaptation are found from
(\ref{nazar1}, \ref{nazar2}) under $\theta<0$ and (\ref{dobro}). We
mention the particular case, where the error probabilities
$\p^1(\bar{\mu}_{2})$ and $\p^2(\bar{\mu}_{1})$ both assume their
minimal values. Consider $|m|\gg 1$ and $b=0$, when
(\ref{dobro}) amounts to
\BEA
\label{golosa}
\left|E_1^1-E_1^2+\frac{\bm_1-\bm_2}{2}    \right|<\frac{\ln 2}{|\theta|}.
\EEA 
The minimal values $\p^1(\bar{\mu}_{2})=\p^2(\bar{\mu}_{1})=1/3$ are
reached when $\left|E_1^1-E_1^2+\frac{\bm_1-\bm_2}{2} \right|\to 0$.
These minimal values are consistent with (\ref{go}). 

Hence the two-state structure is able to adapt imperfectly to two
environmental states that are sufficiently far from each other. 
For the imperfect adaptation we need not only that the temperatures
of $\S$ and $\F$ differ ($\theta\not=1$), but also that the temperature
of $\S$ is {\it negative} ($\theta<0$). 

\subsection{Energy current from $\S$ to $\F$}
\label{energy_disso}

As we have seen, 
adaptation demands that the temperature $\theta$ of $\S$ is different
from that of $\F$ [which is defined to be $1$] and that $\theta<0$. Then there will be a current
of energy $J_{\S\to\F}$ from the bath of $\S$ to that of $\F$. This
energy will flow through $\S+\F$. We now study this energy current
in the stationary regime using (\ref{4}--\ref{usa}).
By definition, the energy flowing from the thermal reservoir of $\S$ is given
as that part of the average energy $\sum_{i\alpha}E^{\alpha}_ip^{\alpha}_i$
change which is driven by the reservoir [see (\ref{4})]:
\BEA
\label{azd}
J_{\S\to\F}=\epsilon
{\sum}_{i\alpha\gamma}E^\alpha_i[\omega^{\alpha\gamma}_{i}p_i^{\gamma}-\omega^{\gamma\alpha}_{i} p_i^{\alpha}].
\EEA
Using the effective master equation [see (\ref{6}, \ref{usa})] we rewrite (\ref{azd}) for the stationary situation as 
\BEA
\label{barb1}
&&\J_{\S\to\F}=\epsilon
{\sum}_{i\alpha\gamma}E^\alpha_i[\omega^{\alpha\gamma}_{i}\p_i^{\gamma}-\omega^{\gamma\alpha}_{i} \p_i^{\alpha}]\\
\label{barb1.5}
&&=-2\epsilon\,\bar{\Omega}^{21}\,\p^1\,\partial_\theta\ln [\,\bar{\Omega}^{21}\bar{\Omega}^{12}\,]
\\
&&=\frac{\epsilon\, e^{\frac{1}{2}(B^{12}_1+B^{12}_2)}}{\bar{\Omega}^{12}+\bar{\Omega}^{21}}\,\,
\frac{m\sinh\left[\frac{(1-\theta)m}{2}  \right]}{2\cosh[\frac{m}{2}]},
\label{barb2}
\EEA
where $m$ is defined in (\ref{sup_polis}). We see that
$\J_{\S\to\F}>0$ for $\theta<1$ confirming that the energy flows from
the lower inverse temperature to the higher one. Thus, $\J_{\S\to\F}>0$
means that the stored energy is lost ({\it dissipated}) in time, i.e.
that the quality of adaptation, which was related to $\theta<0$, tends
to degrade in time. Note that due to the assumed adiabatic limit
$\epsilon\to 0$, $\J_{\S\to\F}>0$ is much smaller than the particle current $\J_\L$.  
The latter is due to the motion of $\F$ and is inversely
proportional to the first power of its characteristic time.

\section{Controller directly sensing two-state environment}
\label{no_cost}

We now show that if we allow the structure $\S$ to interact {\it
directly} with the uncertain environment (reservoirs), there is a
perfect adaptation for a two-state $\S$ having the same temperature
as $\F$. Recall that the temperature of $\F$ was assumed to be $1$, so from now on this is the common temperature of $\S$ and $\F$.
For the present model such an interaction is set
naturally if we assume that the state $\F_i\S^{\a}$ carries $N_i^{\a}$
particles, and it is switched off naturally if $N_i^{\a}$ does depend on
$\a$. Instead of (\ref{5}) we get
\BEA
\label{bdv1}
\omega_i^{\alpha\gamma}=\omega_{i|\L}^{\alpha\gamma}+\omega_{i|\R}^{\alpha\gamma}, \quad i=1,2, \\
\omega_{i|{\rm k}}^{\alpha\gamma}=e^{\frac{1}{2}(E^\gamma_i -E^\alpha_i   )-\frac{\mu_{\rm k}}{2}(N^\gamma_i -N^\alpha_i   )}, \quad
{\rm k}=\L,\R,
\label{bdv2}
\EEA
where we assumed $\theta=1$, since $\S$ and $\F$ have the same temperature [see (\ref{staro})], and we put
$B^{\alpha\gamma}_i=0$ for simplicity. 

The effective transition rates $\bar{\Omega}^{21}$ and $\bar{\Omega}^{12}$ are defined as in (\ref{usa}). They are calculated from (\ref{bdv1}, \ref{bdv2})
by using (\ref{usa}, \ref{7.7}):
\BEA
\label{kaa}
\frac{\bar{\Omega}^{21}}{\bar{\Omega}^{12}}=\frac{\p^2}{\p^{1}}
=e^{E_1^1-E_1^2+\bm(N_1^2-N_1^1)}\,\,
\frac{1+e^{-\Delta E^2+\bm\Delta N^2 }}{1+e^{-\Delta E^1+\bm\Delta N^1 }},
\EEA
where $\Delta E^\a=E^\a_2-E^\a_1$ and $\Delta N^\a=N^\a_2-N^\a_1$. The probabilities $\p^\alpha$ of the states $\S^\a$ are defined via
(\ref{kaa}, \ref{sup7})

A simpler way to obtain (\ref{kaa}) is to note that ${\bar{\Omega}^{21}}/{\bar{\Omega}^{12}}=e^{f^1-f^2}$,
where $f^\a$ is the free energy of the fast $\F$ for a fixed state $\S^\a$ of $\S$:
\BEA
e^{-f^\a}=e^{-E_1^\a+\bm N_1^\a}+e^{-E_2^\a+\bm N_2^\a}
\EEA
Now the state of $\S$ does not depend on the time-scale separation, 
since due to $\theta=1$ the overall state of $\S+\F$  is Gibbsian.

We now assume the uncertain binary environment defined in the previous
section. To ensure that $\F$ functions optimally in each environment we
set $\Delta E^\a=\mu_\a $ for $\a=1,2$ and $\Delta N^1=\Delta
N^2=1$ [cf. with (\ref{mangoost})]. Hence $N^2_i
-N^1_i\equiv N$ does not depend on $i$. 
For the error probabilities we obtain from (\ref{kaa}) [cf. (\ref{dardo4})]:
\BEA
\label{oo1}
\p^2(\bm_1)=\left[ 1+ \frac{2e^{-(E_1^1-E_1^2)-\bm_1N +\bm_2-\bm_1 }}{1+e^{(\bm_2-\bm_1)}} \right]^{-1}, \\
\p^1(\bm_2)=\left[ 1+  \frac{2e^{E_1^1-E_1^2+\bm_2N}}{1+e^{(\bm_2-\bm_1)}}\right]^{-1}.
\label{oo2}
\EEA
Conditions for imperfect adaptation are read-off from 
(\ref{oo1}, \ref{oo2}, \ref{dobro}). There are environments, where 
both error probabilities $\p^2(\bm_1)$ and $\p^1(\bm_2)$ go to zero, e.g. take in  
(\ref{oo1}, \ref{oo2}):
\BEA
\bm_1\gg |\bm_2|=-\bm_2, \quad |N|=-N\gg 1, \quad E_1^1=E_1^2.
\EEA

Thus if $\S$ senses the environment directly, all adaptation
costs disappear: it is possible for $\theta=1$ (equal temperatures for
$\S$ and $\F$, no need for the stored energy), without malfunctions (both
error probabilities can go to zero) and with the minimal number of
states for $\S$. 

\comment{
We also set
\BEA
\label{bekaa}
E_1^1-E_1^2=-\frac{1}{2}(\bm_1+\bm_2)N. 
\EEA
We obtain from (\ref{oo1}--\ref{bekaa}):
\BEA
\label{lao1}
\p^2(\bm_1)=\left[ 1+ \frac{2e^{\frac{1}{2}(\bm_2-\bm_1)N}}{1+e^{(\bm_1-\bm_2)}} \right]^{-1}, \\
\p^1(\bm_2)=\left[ 1+  \frac{2e^{\frac{1}{2}(\bm_2-\bm_1)N}}{1+e^{(\bm_2-\bm_1)}}\right]^{-1}.
\label{lao2}
\EEA
Now both $\p^2(\bm_1)$ and $\p^1(\bm_2)$ go to zero for
$(\bm_2-\bm_1)N\gg 1$. If $\S$ does not couple to the particle
reservoirs, then $N=0$ and (\ref{lao1}, \ref{lao2}) imply that it is
impossible to have simultaneously $\p^2(\bm_1)<1/2$ and
$\p^1(\bm_2)<1/2$.  This is a particular case of the general statement
that no (even imperfect) adaptation is possible if the temperatures
of $\S$ and $\F$ are equal and $\S$ does not sense the environment directly. 
}

\end{document}